\documentclass[12pt,fleqn]{article}

\renewcommand{\baselinestretch}{1.3}

\usepackage{epsfig}
\def\graph#1#2#3{\includegraphics[width=#1, height=#2]{#3}}
\def\graphs#1#2{\includegraphics[scale=#1]{#2}}

\usepackage{epsf,amsfonts,amssymb}

   \def\cD{{\cal D}}
\def\cE{{\cal E}}   
   \def\cL{{\cal L}}
\def\cM{{\cal M}} \def\cN{{\cal N}} \def\cO{{\cal O}} 
\def\cQ{{\cal Q}}   
\def\cU{{\cal U}}

 \def\bn{{\bf n}}

 \def\bB{{\bf B}}  
\def\bE{{\bf E}}

 \def\bR{{\bf R}} \def\bS{{\bf S}} \def\bT{{\bf T}}
   
 \def\bZ{{\bf Z}}

\def\Nbar{{\overline N}}

\def\nbar{{\overline n}}

\def\veck{{\vec{k}}}
\def\vecm{{\vec{m}}}
\def\vece{{\vec{e}}}

\def\beq#1{\begin{equation}\label{#1}}
\def\eeq{\end{equation}}
\def\beqa#1{\begin{eqnarray}\label{#1}}
\def\eeqa{\end{eqnarray}}
\def\bfig#1#2{\begin{figure}[#1]\label{#2}}
\def\efig{\end{figure}}
\def\bcen{\begin{center}}
\def\ecen{\end{center}}

\def\bra{\left\langle}
\def\ket{\right\rangle}
\def\tr{{\rm tr}\,}

\makeatletter \@addtoreset{equation}{section} \makeatother

\title{ {\small \begin{flushright}  IFT UAM/CSIC-2005-56\\ {\tt hep-th/0512303}\end{flushright}} \ \\{\bf \Large Large N orbifold field theories on the twisted torus}}

\date{}

\author{\textsc{\normalsize C. Hoyos}\\ \\{\small \sl  Instituto de F\'{\i}sica Te\'orica UAM/CSIC,} \\{\small \sl  Facultad de Ciencias, Cantoblanco. 28049 Madrid, Spain }\\{\small \tt c.hoyos@uam.es}}

\begin{document}
\renewcommand{\baselinestretch}{1.0}
\maketitle
\renewcommand{\baselinestretch}{1.3}

\begin{abstract}
We study the planar equivalence of orbifold field theories on a small three-torus with twisted boundary conditions, generalizing the analysis of hep-th/0507267. The nonsupersymmetric orbifold models exhibit different large N dynamics from their supersymmetric "parent" counterparts. In particular, a moduli space of Abelian zero modes is lifted by an O($N^2$) potential in the "daughter" theories.  We also find disagreement between the number of discrete vacua of both theories, due to fermionic zero modes in the parent theory, as well as the values of semiclassical tunneling contributions to fermionic correlation functions, induced by fractional instantons.
\end{abstract}

\vskip7.0cm
\begin{flushleft} February 2006 \end{flushleft}
\vfill
\addtocounter{page}{-1}
\thispagestyle{empty}

\newpage

\section{Introduction.}

Although there have been considerable advances in the strong-coupling description of gauge theories since their formulation, this remains an unresolved matter.

A line of attack to this problem has been to analyze related systems with simpler dynamics. Specially successful has been the study of supersymmetric theories, that in some cases can be solved exactly. However, the applicability of the methods employed usually relay on supersymmetry and cannot be used safely for non-supersymmetric theories. A different and older approach is 't Hooft's large $N$ expansion \cite{ref:thooftlargen}, where each term in the expansion is non-perturbative in the coupling constant. Although it gives a good qualitative description of many features of strongly coupled gauge theories, extracting quantitative results for realistic theories has been demonstrated to be a difficult task.

Recently, a new combination of both lines has been proposed under the name of planar equivalence \cite{ref:strassler}. Starting from a supersymmetric theory, called 'parent', it is projected to a non-supersymmetric theory, called 'daughter', so that they are equivalent in the large $N$ limit. The projections are inspired in string constructions and they are named after orbifolds \cite{ref:orbifold} and orientifolds \cite{ref:orientifold}. Orbifold projection relates a $\cN=1$ $SU(kN)$ supersymmetric gauge theory with a $(SU(N))^k$ gauge theory with fermions in bifundamental representations under two adjacent gauge group factors. Orientifold projection relates a $\cN=1$ $SU(N)$ supersymmetric gauge theory with a $SU(N)$ gauge theory with fermions in two-index symmetric or antisymmetric representations. For these theories, planar equivalence has been proved at the perturbative level, and there are formal proofs of exact equivalence \cite{ref:largenequiv,ref:orientifoldproof}. A remarkable example of this program is the computation of the quark condensate in terms of $\cN=1$ gaugino condensate using orientifold theories \cite{ref:quarkcond}. Further works on finite $N$ computations in orientifold theories using string duals can be found in \cite{ref:newresults1}. Theories with other possible representations of fermions and their planar properties are studied in \cite{ref:newresults2}.

Although the results are encouraging, the exact equivalence is not completely established, specially for orbifold theories, that have been the subject of much discussion \cite{ref:forandagainst,ref:yaffetheta,ref:shifmanlast,ref:tong,ref:ourpaper}, although there are also some concerns about orientifold theories \cite{ref:ourpaper}. We want to clarify this point by studying these theories in a dynamically controlled regime, where we can analyze the behavior of both parent and daughter theory and compare them. In order to do that, we will introduce them in a small spatial torus $\bR\times \bT^3$, where the theory is in a weak coupling regime, so perturbative and semiclassical methods can be used. The first analysis of planar equivalence using finite volume is \cite{ref:tong}, where a single spatial direction was compactified in a circle. It was shown that the dynamics of the orbifold daughter breaks down planar equivalence for small enough radius, because the sector where it would hold becomes tachyonic. However, it is not clear that this will remain true when the circle is decompactified. The extension to a total compactification of space was made in \cite{ref:ourpaper}. In this case non-trivial flat connections in the torus are zero modes that generate a moduli space. In the small volume limit, zero modes control the dynamics of the theory, so we can concentrate on the study of the moduli space. In principle, quantum corrections can generate a potential of order $O(N^2)$ over the moduli space, but supersymmetry guarantees that a potential is not generated. Planar equivalence boils down to the vanishing of the daughter effective potential to $O(N^2)$, but it is shown that this is violated locally at some points of the moduli space due to non-linear effects, although it is satisfied on the average. Also, the $O(N)$ potential makes the physics of parent and daughter moduli space very different. It seems that the signals of a possible failure of planar equivalence are a consequence of having scalar degrees of freedom that arise when we compactify the theory, and that their effects may disappear when we go to the infinite volume limit. Then, it is interesting to get rid of the moduli space in order to avoid this problem. This can be done using twisted boundary conditions, that will constitute the frame of this paper. 

Orbifold theories are invariant under a global $\bZ_N$ group of symmetry, the center, that can be used to introduce twisted boundary conditions and lift most or all of the non-Abelian moduli space. Orientifold theories are invariant only under a $\bZ_2$ center group, that is not enough to significantly change the moduli space. For this reason, this paper will concentrate on orbifold theories. Twisted boundary conditions will allow the construction of a precise mapping of topological sectors between parent and daughter, so we can compare the vacuum sectors. Moreover, we will be able to compare fermionic correlation functions induced by tunneling between vacua in each theory. Ordinary instantons can contribute to such quantities, but they are not useful to answer questions about planar equivalence because their semiclassical contributions vanish exponentially in the $N\to \infty$ limit. The reason is that their action scales as $S_{\rm inst}=8\pi^2N/g_t$, where $g_t=g_{YM}^2N$ is the 't Hooft coupling. However, when twisted boundary conditions are introduced, new vacua and configurations associated to tunneling appear. This new class of tunneling configurations can give a non-vanishing contribution to fermionic correlation functions, since their action scales as a fractional instanton $S_{\rm tun}=8\pi^2/g_t$.

The outline of the paper is as follows. In section 2 we introduce twisted boundary conditions and show how they can be used to make a complete map between different gauge bundles of parent and daughter theory. In  section 3 we use this map to describe and compare vacuum states of both theories. In section 4 we extend the analysis to include effects induced by tunneling. Finally, we give a geometrical interpretation in terms of bound states of D-branes and conclude.

\section{Embedding of the orbifold theory.}

We are interested in finding the map between physical configurations of parent and daughter theories relevant for planar equivalence. According to \cite{ref:largenequiv}, we should compare the vacuum sectors of both theories. However, this is a difficult task, specially for the daughter theory, because we cannot use perturbation theory and we do not have non-perturbative effects under control in the IR limit. In order to avoid these problems, we introduce parent and daughter theories in a small finite volume. This has been an extensively used tool to study asymptotically free gauge theories, since they are in a weak coupling regime where perturbative and semi-classical methods are applicable.

The orbifold theories that we are considering have a $(U(N))^k$ gauge group and Weyl fermions that transform as the bifundamental representation $(N_i,\Nbar_{i+1})$ of two adjacent gauge groups. These theories can be obtained from a $\cN=1$ $U(kN)$ supersymmetric gauge theory after performing an ``orbifold projection''.
\beq{eq:orbifoldproj}
A_\mu\to \left( \begin{array}{cccc} A_\mu^1 & & & \\ & A_\mu^2 & & \\& &\ddots & \\ & & & A_\mu^k \end{array}\right) \ \ ; \ \ \lambda \to \left( \begin{array}{cccc}  0& \lambda_{1,2} & & \\ & \ddots & \ddots & \\ & & 0& \lambda_{k-1,k}\\\lambda_{k,1} & & &0 \end{array}\right)
\eeq
in general, when summing over different orbifold components $A_\mu^i$, $\lambda_{i,i+1}$, we will always assume that indices are defined modulo $k$. For this kind of theories, it has been proved that parent and daughter planar diagrams give the same result, if the coupling constants of both theories are related by $kg_p^2=g_d^2$.
 
We should point that the gauge group of the daughter theory is a subgroup of the gauge group of the parent theory {\em of the same rank}. This is quite useful if we try to establish a map between both theories. For instance, every representation of the parent theory has a unique decomposition in representations of the daughter theory. Notice that if we ignore Abelian groups in the daughter theory, the rank would be different. We will see that Abelian groups play a relevant role in the mapping. We will ignore only the diagonal $U(1)$ group of the daughter theory, that maps trivially to the Abelian part of the parent group. All fields are uncharged under these groups, so they decouple trivially.

\subsection{Twisted boundary conditions in the torus.}

Twisted boundary conditions were introduced by 't Hooft \cite{ref:emflux} and successfully applied by Witten to study supersymmetric theories \cite{ref:index,ref:index2}. The topics we review briefly here can be found more thoroughly studied in these references. An extensive classical analysis of twisted boundary conditions in $\bT^4$ can also be found in \cite{ref:ymtorus}.

When we compactify a gauge theory in a three-torus $\bR\times \bT^3$, we have some freedom to specify the boundary conditions of the fields on the sides of the torus. For a $SU(M)$ gauge group, the fields must be periodic up to gauge transformations. In order to be in a configuration with zero field strength, the gauge transformations on each side must commute. We can always change the boundary conditions by making gauge transformations and redefining our field, so all these configurations are equivalent. In particular, all can be reduced to the case with trivial periodic boundary conditions, where there are zero modes for the gauge fields given by constant Abelian connections. 

For the theories we are interested, the fields are invariant under a discrete subgroup of the gauge group, that we call the invariant center. For instance, for a (supersymmetric) $SU(M)$ gauge theory, the center is $\bZ_M$. As a consequence, we can consider our theory as a $SU(M)/\bZ_M$ theory and impose non-trivial boundary conditions on the fields. The fields are periodic up to a gauge transformation, but now the transformations on each side need to commute only up to an element of the center. Therefore, for each side we can give an element of the center that defines non-equivalent boundary conditions. The information is encoded in the so-called magnetic flux $\vecm$, which is a three-vector of integers defined modulo $M$. Formally, we are constructing gauge bundles of different topology, using the mapping between non-trivial one-cycles of the group $SU(M)/\bZ_M$ to the torus. Since in general we cannot reduce these configurations to the trivial case, there are no zero modes for gauge fields.  

Imagine that we are in a sector of definite magnetic flux, and we choose to work in the $A_0=0$ gauge. This gauge only fixes time-dependent transformations. When we make a time-independent gauge transformation, it has to satisfy the same boundary conditions as the fields, up to an element of the center. Then, we can introduce an element of the center for each direction that characterizes the twisted gauge transformation. We can group them in a single three-vector $\veck$ of integers. Twisted gauge transformations are not continuously connected to the identity, so physical states do not need to be invariant under them. As a matter of fact, each time we make a twisted gauge transformation we can reach a different sector of the Hilbert space. Using Fourier transformation, we can build states invariant under the action of a twisted gauge transformation, labelled by the electric flux $\vece$, a three-vector of integers defined modulo $M$.

The information carried by electric and magnetic flux can be encoded in a single quantity $n_{\mu\nu}$, the {\em twist tensor}, that emerges when we study gauge bundles in an Euclidean $\bT^4$. As in $\bT^3$, we use gauge transformations to glue the sides of the torus. The gauge transformations on the sides $\mu$ and $\nu$ must commute up to an element of the center given by $\exp(2\pi i n_{\mu\nu}/M)$.
\beq{eq:consistency}
\Omega(\mu,x+{\bf l_\nu})\Omega(\nu,x) = \exp\left({2\pi i\over M} n_{\mu\nu}\right)\Omega(\nu,x+{\bf l_\mu})\Omega(\mu,x)
\eeq

The twist tensor is an antisymmetric tensor of integers defined modulo $M$. It is related to electric and magnetic flux in a similar way as electric and magnetic components are related to field strength 
\beq{eq:fluxes}
\begin{array}{rcl}
n_{ij}  & = & \epsilon_{ijk} m_k \\
n_{0i} & = & k_i 
\end{array}
\eeq
In general, the Euclidean gauge bundles will have a curvature. In this case, the action does not vanish, but it has a minimum value given by the Pontryagin number or topological charge. Usually, contributions from ordinary instantons are considered. They come from bundles associated to the wrapping of the gauge group over a $S^3$. Then, this contribution is an integer known as the winding number. For gauge bundles associated to the torus, the contribution to topological charge can be fractional, because we are using a representation that is not faithful for $SU(M)/\bZ_M$. In the sector of zero winding number, the topological charge is given by  
\beq{eq:topcharge}
\cQ  =  {1\over 16\pi^2}\int_{\bT^4} \tr F_{\mu\nu}(x){\tilde F}_{\mu\nu}(x)d^4 x =  -{1\over 4M} n_{\mu\nu} {\tilde n}_{\mu\nu}=  - {1\over M} \veck \cdot \vecm
\eeq
where ${\tilde A}_{\mu\nu}=(1/2)\epsilon_{\mu\nu\rho\sigma}A_{\rho\sigma}$. 

\subsection{Twisted boundary conditions and orbifold theories.}\label{sec:map}

The parent theory is a $SU(M)$ supersymmetric gauge theory with $M=kN$. All the previous discussion can be applied directly to this case. In order to study the relation between parent and daughter theories in this setup, we follow the construction that 't Hooft used to find self-dual solutions of fractional topological charge \cite{ref:torons}, called torons, and generalize it to our case. The analysis is made in the Euclidean $\bT^4$.

We split rows and columns in $k$ diagonal boxes of $N\times N$ size. Then, we work in the subgroup $(SU(N))^k\times U(1)^{k-1}\subset SU(kN)$, which is the orbifold projection on the gauge sector. If we had considered only a $(SU(N))^k$ theory, without the Abelian part, then the possible bundles in the daughter theory would have been characterized by a $\bZ_N$ diagonal subgroup of $(\bZ_N)^k$. This is so because of the fermions in the bifundamental representations. If twist tensors of different $SU(N)$ groups were different, boundary conditions for fermions will not be consistent, according to (\ref{eq:consistency}). However, Abelian fields can be used to absorb extra phases and enlarge the number of possible bundles. This will be evident in our construction.

Let $\omega_i$, $i=1,\dots, k$ be the $U(1)$ generators
\beq{eq:u1generators}
\omega_i=2\pi {\rm diag}\,\left( -N{\bf 1}_N, \cdots , N(k-1){\bf 1}_N, \cdots, -N{\bf 1}_N\right)  
\eeq
where $N(k-1){\bf 1}_N$ is located at the $i$th position and ${\bf 1}_N$ denotes the $N\times N$ identity matrix. Note that only $k-1$ generators are independent, since $\sum_{i=1}^k \omega_i =0$.

Define the following twist matrices
\beq{eq:twistmatrices}
V_iW_i=W_iV_i\exp\left({2\pi i\over kN}+{i\omega_i\over kN^2}\right)
\eeq
All other pairs commute. Notice that although $V_i$ and $W_i$ are matrices defined in principle to make a $SU(kN)$ twist, in fact the terms in the exponent are such that the non-trivial twist is made only over the $i$th $N\times N$ box. A possible representation is
\beq{eq:vwrep}
\begin{array}{lcl}
V_i & = & {\rm diag}\,\left( {\bf 1}_N,\cdots , P_N^{(i)},\cdots, {\bf 1}_N  \right)\\
W_i & = & {\rm diag}\,\left( {\bf 1}_N,\cdots , Q_N^{(i)},\cdots, {\bf 1}_N  \right)
\end{array}
\eeq
where $Q_N^{(i)}$ and $P_N^{(i)}$ are the $SU(N)$ matrices of maximal twist $P_NQ_N=Q_NP_Ne^{2\pi i/N}$ associated to the $i$th gauge group.

Let us try the ansatz
\beq{eq:ansatz}
\Omega(\mu,x)=\left(\prod_{i=1}^k P_i^{a_\mu^i}Q_i^{b_\mu^i}\right)\exp\left(i\sum_{i=1}^k \omega_i \alpha_{\mu\nu}^i x_\nu/l_\nu\right)
\eeq
where summation over $\nu$ is understood. The numbers $a_\mu^i$, $b_\mu^i$ are arbitrary integers. $\alpha_{\mu\nu}^i$ is an antisymmetric real tensor which will be associated to the Abelian fluxes that are necessary to cancel out phases in the case of non-diagonal twist.

Inserting (\ref{eq:ansatz}) in the consistency conditions (\ref{eq:consistency}) we find that
\beq{eq:twisttensors}
\begin{array}{rcl}
n_{\mu\nu} & = & \sum_{i=1}^k n_{\mu\nu}^i\\
n_{\mu\nu}^i & = & a_\mu^i b_\nu^i- a_\nu^i b_\mu^i
\end{array}
\eeq
and the $k$ conditions
\beq{eq:conditions}
\sum_{j=1}^k n_{\mu\nu}^j- kn_{\mu\nu}^i= 2kN^2\left( k\alpha_{\mu\nu}^i - \sum_{j=1}^k \alpha_{\mu\nu}^j \right) \ \ , \ \ \sum_{i=1}^k \alpha_{\mu\nu}^i=0
\eeq
which can be used to obtain the $\alpha_{\mu\nu}^i$ tensors from the $n_{\mu\nu}^i$ tensors. The second equation in (\ref{eq:conditions}) comes from the fact that there are only $k-1$ independent $U(1)$ groups. Any other linear condition will be valid.

This establishes a map between Euclidean bundles in parent and daughter theories, identifying $n_{\mu\nu}^i$ as the twist tensors of the daughter theory. This implies that many inequivalent twists in the daughter theory map to the same twist in the parent. A second consequence relies on the fact that $\alpha_{\mu\nu}^i$ are constant Abelian field strengths. This implies that configurations like 't Hooft's torons in the parent theory can be mapped to Abelian electric and magnetic fluxes in the daughter, showing that Abelian groups play a relevant role in the mapping. Generically, physically inequivalent configurations in the daughter theory, like Abelian fluxes and torons, map to the same configuration (up to gauge transformations) in the parent theory. The map can be one-to-many, and it is given by the possible decompositions of the parent twist tensor $n_{\mu\nu}$ (defined mod $kN$) in $k$ daughter's twist tensors $n_{\mu\nu}^i$ (each defined mod $N$).

A remark is in order. Although in principle we can turn on arbitrary fluxes of Abelian fields in the daughter theory, we must take into account the presence of charged fermions under these. Since bifundamentals correspond to off-diagonal boxes in a $SU(kN)$ matrix, the value of Abelian fluxes is determined modulo $N$ by consistency conditions of the twist. This implies that fractional contributions of Abelian fluxes to the topological charge are determined completely by (\ref{eq:conditions}). On the other hand, integer contributions to the topological charge in the parent theory can map to non-Abelian instantonic configurations, Abelian fluxes of order $N$ or a mixture of both. Since the energy of this kind of configurations is a factor $N$ larger than fractional contributions, they are irrelevant in the large $N$ limit and we will not worry about them anymore.

We can be more precise with the map of Abelian configurations. The $U(1)$ groups under consideration follow from the decomposition of $SU(kN)$ gauge connections in boxes $A_\mu = \,{\rm diag}\,(A_\mu^1,\cdots ,A_\mu^N)$ so that $U(1)$ connections are given by 
\beq{eq:theu1s}
B_\mu^i = \tr A_\mu^i - \tr A_\mu^{i+1} 
\eeq
where $i=1,\dots,k-1$. Then, following \cite{ref:cita27toni}, we can relate topological invariants of Abelian groups of the daughter theory to the twist tensor of the parent theory. Let $F_{\mu\nu}^i$ be the field strength of the $i$th Abelian group. We have the following topological invariants 
\beq{eq:chern1}
c_{\mu\nu}^i={1\over 2\pi}\int dx_\mu \wedge dx_\nu \tr F_{\mu\nu}^i,
\eeq
given by the integral of the first Chern class over non-contractible surfaces of the torus and 
\beq{eq:chern2}
\cQ^i={1\over 16\pi^2}\int_{\bT^4} \tr (F_{\mu\nu}^i {\tilde{F}}_{\mu\nu}^i) d^4 x
\eeq
which is the Pontryagin number and it is related to first and second Chern classes. We can relate $\cQ^i$ to the topological charge of the parent theory, while $c_{\mu\nu}^i$ (mod $N$) will be a twist tensor $n_{\mu\nu}'$ for the $SU(N)\subset SU(kN)$ subgroup of gauge connection $A_\mu^i-A_\mu^{i+1}-B_\mu^i$. It contributes to a fractional topological charge through (\ref{eq:topcharge}).

\section{Vacuum structure.}

We have been able to construct a map between configurations of both theories, showing that they are physically inequivalent, although this is not a surprise. As a matter of fact, we are interested only in large $N$ equivalence. According to \cite{ref:largenequiv}, planar equivalence will hold non-perturbatively in the sector of unbroken $\bZ_k$ orbifold symmetry. The necessary and sufficient condition is that operators related by orbifold projection, and invariant under orbifold transformations, have the same vacuum expectation values. Therefore, it is enough to check the vacuum sector. In infinite volume it is argued that large $N$ equivalence does not hold for $k>2$ because of spontaneous breaking of orbifold symmetry by the formation of a gaugino condensate in the parent theory \cite{ref:yaffetheta}. For $k=2$ the condensate does not break orbifold symmetry and the question remains open. In finite volume in principle there is no spontaneous breaking, so any failure of large $N$ orbifold equivalence must show itself in a different way. In the following analysis we will restrict to the sector of vanishing vacuum angles for simplicity.

\subsection{Ground states in the daughter theory.}

We will start by comparing the bosonic sector of ground states of parent and daughter theories in a three-torus. The first question we can ask ourselves is what are the relevant configurations we must consider. We assume that the orbifold projection has been made at a scale of energy much larger than the typical scale of the torus $\mu >> 1/l$. At this point, all the coupling constants of the orbifold theory are equal. The running of coupling constants of non-Abelian groups is given at leading order by the renormalization group of the parent theory, by perturbative planar equivalence. However, the running for Abelian groups is different since they are not asymptotically free. As a matter of fact, the coupling constants of Abelian groups will be smaller than the coupling constants of non-Abelian groups in our torus, and the difference will increase with the volume. We can now consider what are the contributions to energy of Abelian configurations, that depend on electric ($E_i=\alpha_{0i}$) and magnetic ($B_i=\epsilon_{ijk}\alpha_{jk}$) fluxes as 
\beq{eq:hamabelian}
H=\int_{\bT^3_l}\left({e^2(l)\over 2} \bE^2+{1\over 2e^2(l)}\bB^2\right)
\eeq
In the quantized theory $\bB^2$ acts as a potential, while $\bE^2$
plays the role of kinetic energy. 

We see that the magnetic contribution is proportional to the inverse of the coupling, so the ground states of the theory will be in the sectors of zero Abelian magnetic flux. From (\ref{eq:twisttensors}), this forces us to introduce equal magnetic flux on all non-Abelian gauge factors, 
\beq{eq:orbftwist}
\vecm_p  =  k\vecm_d
\eeq
where $\vecm_{p(d)}$ is the magnetic flux in the parent (daughter) theory. This kind of boundary conditions preserve the orbifold symmetry of interchanging of $SU(N)$ groups. 

The twist of the daughter theory is determined by a $\bZ_N$ diagonal subgroup. The elements of this group are actually equal to the elements of a $\bZ_N$ subgroup of the center of the parent theory
\beq{eq:centerelements}
Z_l=\bigoplus_k \,{\rm diag}\,\left(1, e^{2\pi il/N}, \dots, e^{2\pi i l(N-1)/N} \right)
\eeq
We are now interested in imposing twisted boundary conditions in the daughter theory such that the moduli space of non-Abelian gauge groups is maximally lifted. We can introduce a unit of magnetic flux in the same direction for each of the gauge groups, so we are performing a maximal twist of the theory and thus lifting all the non-Abelian moduli space. However, the kind of boundary conditions we are imposing does not prevent the existence of constant connections for Abelian groups. Bifundamental fermions are charged under these $U(1)$ groups, so it is not possible to gauge away constant connections. Moreover, they become periodic variables. Therefore, the Abelian moduli space of the daughter theory is a ($k-1$)-torus (for each spatial direction).

In the $A_0=0$ gauge we have freedom to make gauge transformations depending on spatial coordinates. Abelian gauge transformations must be such that they compensate the phases that can appear on fermionic fields if we make twisted gauge transformations that are not equal for all non-Abelian gauge groups. Only transformations with $\veck\cdot \vecm \neq 0$ mod $N$ lead to a different vacuum, so there are $N^k$ possible non-Abelian vacua. The total moduli space consists on $N^k$ disconnected $\bT^{3(k-1)}$ tori. The Abelian part of the vacuum is a wavefunction over the torus. 

In our setup, orbifold symmetry is preserved in the vacua that are reached by twisted gauge transformations of the same topological class for all non-Abelian gauge groups. For these, Abelian gauge transformations are trivial. In Sec.~\ref{sec:moduli} we will see that these vacua are dynamically selected by the physics on Abelian moduli space.

\subsection{The mapping of vacua.}

We would like to study planar equivalence in a physical setup that is equivalent for the parent theory. From (\ref{eq:orbftwist}) this corresponds to the introduction of $k$ units of magnetic flux in the same direction as in the daughter theory. In this case we can reach $N$ disconnected sectors making twisted gauge transformations. We would like to see how these vacua can be mapped to the orbifold-preserving vacua of the daughter theory. In order to do that, we should be able to construct the operators associated to twisted gauge transformations in the daughter theory from the corresponding ones in the parent theory. A second condition is that we should be able to identify the moduli spaces of both theories.

Let us examine the first condition. Suppose that we introduce $k$ units of magnetic flux in the $x_3$ direction in the parent theory: 
\beq{eq:boundcond}
\begin{array}{lcl}
A_\mu(x_1+L,x_2,x_3) & = & {\tilde V}_kA_\mu(x_1,x_2,x_3){\tilde V}_k^{-1} \\
A_\mu(x_1,x_2+L,x_3)  & = & {\tilde W}_kA_\mu(x_1,x_2,x_3){\tilde W}_k^{-1} \\
A_\mu(x_1,x_2,x_3+L) & = & A_\mu(x_1,x_2,x_3)
\end{array}
\eeq
where twist matrices must satisfy ${\tilde V}_k{\tilde W}_k = {\tilde W}_k{\tilde V}_ke^{2\pi i k/kN}$. We can go from one vacuum state to other by making a twisted gauge transformation $U$, that satisfies the above conditions (\ref{eq:boundcond}), except in the $x_3$ direction, where
\beq{eq:twistgaugecond}
U(x_1,x_2,x_3+L)=e^{2\pi i k/kN}U(x_1,x_2,x_3).
\eeq
A good election for the twist matrices is ${\tilde V}_k={\bf 1}_k\otimes P_N$, ${\tilde W}_k = {\bf 1}_k\otimes Q_N$. Moreover, we can use the $SU(k)$ group that is not broken by the boundary conditions to rewrite $U$ as a box-diagonal matrix, that can be mapped to the twisted gauge transformations of the daughter theory, if we arrange them in a single matrix. In general, the transformations in the daughter theory will include Abelian phases, so the map of vacua will be from one in the parent to many in the daughter. 

We now turn to the second condition. The twisting is not enough to lift completely the moduli space of the parent theory, but a torus $\bT^{3(k-1)}$ remains. As a matter of fact, it is an orbifold, since we should mod out by the Weyl group of $SU(k)$. In the daughter theory, the Abelian moduli space is the same, except that there are no Weyl symmetries acting on this space. However, in the daughter theory there are discrete global symmetries associated to the permutations of factor groups that can be mapped to Weyl symmetries of the parent theory. They do not modify the moduli space, but we can make a projection to the invariant sector of the Hilbert space to study planar equivalence.
  
There will be a wavefunction over the moduli space associated to the vacuum. In the parent theory, supersymmetry implies that the wavefunction is constant. In the daughter theory there is no supersymmetry, so a potential can be generated that will concentrate the wavefunction at the minima. Another consequence of supersymmetry is that there are fermionic zero modes over the moduli space, which give rise to more zero-energy states. We will investigate both questions in the next sections.
\subsection{Electric fluxes and vacuum angles.}\label{sec:elecflux}

The vacuum states we have studied in the parent theory are generated from the trivial vacuum $\left| 0_p \ket$, associated to the configuration of magnetic flux $\vecm=(0,0,k)$ and classical gauge connection $A_\mu=0$, by operators that implement the minimal twisted gauge transformations on the parent Hilbert space $ {\hat U}_p$, characterized by $\veck=(0,0,-1))$ as the electric component of the twist tensor (\ref{eq:fluxes}). 
\beq{eq:kstates}
\left| l \ket = {\hat U}_p^l\left|0_p \ket 
\eeq
We can build Fourier transformed states of (\ref{eq:kstates}) that only pick up a phase when we make a twisted gauge transformation. We must take into account that the operators associated to twisted transformations depend on the vacuum angle $\theta_p$, so we restrict to a sector of the Hilbert space with definite vacuum angle. Then,
\beq{eq:windingone}
{\hat U}_p^N = e^{i\theta_p}{\bf \hat 1}
\eeq
The states we are considering are labelled by the electric flux $\vece=(0,0,e)$%
\beq{eq:estates}
\left|e\ket = {1\over \sqrt{N}} \sum_{l=0}^{N-1} e^{-{2\pi il\over kN}\left(e+{\theta_p\over 2\pi}k \right)}  {\hat U}_p^l\left|0_p\ket
\eeq
and transform as
\beq{eq:transfstate}
{\hat U}_p\left|e\ket= e^{{2\pi i\over kN}\left(e+{\theta_p\over 2\pi}k \right)}\left|e\ket
\eeq
Notice that $e$ must be a multiple of $k$, in order to be consistent with (\ref{eq:windingone}).

The theory must be invariant under $2\pi$ rotations of the vacuum angle $\theta_p\to \theta_p+2\pi$. The states (\ref{eq:estates}) are not invariant under this transformation, but it generates a spectral flow that moves a state to another one $e\to e+k$. In our particular case we can define a $e=ke_p$ electric flux such that the spectral flow acts as $e_p\to e_p+1$ 
\beq{eq:ekstates}
\left|e_p\ket = {1\over \sqrt{N}} \sum_{l=0}^{N-1} e^{-{2\pi il\over N}\left(e_p+{\theta_p\over 2\pi} \right)}  {\hat U}_p^l\left|0_p\ket
\eeq

In the daughter theory, the vacua are obtained from the trivial vacuum $\left|0_d \ket$, associated to magnetic fluxes $\vecm^i=(0,0,1)$ and zero classical gauge connections $A_\mu^i=0$, by the set of operators that implement minimal twisted gauge transformations on the daughter Hilbert space ${\hat U}_i$, characterized by $\veck^i=(0,0,-1)$ 
\beq{eq:lstates}
\left|l_1,\cdots,l_k \ket = \prod_{i=1}^k{\hat U}_i^{l_i} \left|0_d \ket
\eeq 
We will concentrate on states that are reached by diagonal twisted transformations, so that they are characterized by a single integer $l_i=l$, $i=1,\dots,k$. Again, we can build the Fourier transformed states, taking into account that we are in a sector of the Hilbert space with definite vacuum angles $\theta_i$, so
\beq{eq:windingoness}
{\hat U}_i^N=e^{i\theta_i}{\bf \hat 1}
\eeq
The states are labelled by electric fluxes $e_i$
\beq{eq:elstates}
\left|e_1,\cdots,e_k\ket_D ={1\over \sqrt{N}}\sum_{l=0}^{N-1} e^{-{2\pi il\over N}\left(\sum_i e_i+{\sum_i \theta_i\over 2\pi} \right)}  \prod_{i=1}^k{\hat U}_i^l\left|0_d\ket  
\eeq
If we had considered transformations other than diagonal, this would be shown in the exponent, where different combinations of electric fluxes and vacuum angles will appear. If we make a diagonal twisted transformation, these states change as
\beq{eq:transfdiagstate}
\prod_{i=1}^k{\hat U}_i\left|e_1,\cdots,e_k\ket_D= e^{{2\pi i\over N}\left(\sum_i e_i+{\sum_i \theta_i\over 2\pi} \right)}\left|e_1,\cdots,e_k\ket_D
\eeq
If the twisted transformation are non-diagonal, there are extra phases appearing on the boundary conditions of the fields that must be compensated by Abelian transformations, so the state will change to a different vacuum in the Abelian sector. We can define subsectors of the Hilbert space that are invariant under diagonal transformations and that are shifted to other sectors by non-diagonal transformations. The labels for such sectors will be given by the Abelian part, so there will be $N^{k-1}$ of them.
%Staying in the sector of diagonal states, the spectral flow induced by rotating {\em one} of the vacuum angles $\theta_i\to \theta_i+2\pi$, shifts the sum of electric fluxes $\sum_i e_i \to \sum_i e_i +1$. It is tempting then to define a diagonal electric flux $e_D=\sum_i e_i$ and a diagonal vacuum angle $\theta_D=\sum_i \theta_i$, that will be mapped in the parent theory to the electric flux $e_p$ and the vacuum angle $\theta_p$. Notice that this is {\em different} to the usual mapping of vacuum angles, where it is assumed that all are equal $\theta_i=\theta_d$, so that we map $\theta_p=k\theta_d$. The difference consists in that $\theta_d$ is seen as a vacuum angle of the daughter theory. However, rotating $\theta_d$ implies a rotation of all the vacuum angles in the daughter theory, and rotating $\theta_d$ by $2\pi/k$ is physically different from rotating a single $\theta_i$ by $2\pi$, even if both will rotate $\theta_D$ by $2\pi$. 

%In other words, the $k$ spectral flows induced by each of the vacuum angles of the daughter theory $\theta_i$, map to the single spectral flow of the vacuum angle of the parent theory $\theta_p$, when we restrict to the sector of diagonal states. This analysis suggests that, contrary to the proposal made in \cite{ref:yaffetheta}, we should not associate the spectral flow of the parent vacuum angle $\theta_p$ to the "spectral flow "of the daughter vacuum angle $\theta_d$, that has no physical meaning in the daughter theory.

\subsection{Fermionic zero modes.}

Fermionic zero modes are spatially constant modes that satisfy the Dirac equation with zero eigenvalues. In the parent theory the gaugino has $k-1$ zero modes
\beq{eq:zeromodes}
{\bf \lambda_0}^a = {\bf \lambda}_{k\times k}^a \otimes {\bf 1}_N 
\eeq
where $a=1,\dots, k-1$ runs over the Cartan subalgebra of the unlifted $SU(k)$ subgroup. If we go to a Hamiltonian formulation, as in \cite{ref:index}, they do not contribute to the energy. We can define  creation and annihilation operators for these modes:  ${\bf \it a}_a^{*\,\alpha}$ and  ${\bf \it a}_a^{\alpha}$ where $\alpha=1,2$ refers to spin and $a=1,\dots, k-1$ is the gauge index. Physical states must be gauge invariant. For zero modes gauge invariance appears in the form of Weyl invariance, a discrete symmetry that stands after fixing completely the gauge. The set of Weyl-invariant operators that we can construct with creation operators is limited to contract the gauge indices with $\delta_{ab}$. Thus, the only allowed operators that create fermions over the gauge vacuum are powers of 
\beq{eq:creation}
\cU = \epsilon_{\alpha\beta}{\bf \it a}_a^{*\,\alpha} {\bf \it a}_a^{*\,\beta} 
\eeq
Fermi exclusion principle limits the number of times we can apply $\cU$ over the vacuum, so the set of possible vacua is
\beq{eq:fermionvacua} 
\left|0_B\ket,\ \cU\left|0_B\ket, \ \dots, \ \cU^{k-1}\left|0_B\ket.
\eeq
Some remarks are in order. First, the total number of vacua in the parent theory is $kN$, in agreement with the Witten index. Second, $\cU$ acquires a phase under general chiral transformations, so the chiral symmetry breaking vacua of infinite volume might be constructed from appropriate linear combinations of the states (\ref{eq:fermionvacua}).

What is the situation in the daughter theory? The only possible candidates to give fermionic zero modes in spite of twisted boundary conditions are the trace part of bifundamental fields. However, those fields are charged under Abelian groups, so it is not possible to find zero modes over the whole moduli space. We are then confronted to the same kind of concern as in \cite{ref:yaffetheta}, the number of vacuum states is different in each theory, even in the sector of vacua that preserve orbifold symmetry.

A necessary condition for planar equivalence to hold is that the vacuum expectation values of orbifold invariant quantities are the same up to factors of $k$ given by the orbifold mapping \cite{ref:largenequiv}. In infinite volume, this could happen only for $k=2$. Otherwise, the vacua of the parent theory spontaneously break orbifold symmetry \cite{ref:yaffetheta}. However, in the $\bZ_2$ orbifold the number of parent vacua is twice the number of daughter vacua so planar equivalence would be true only if there is a two-to-one mapping between vacua of parent and daughter theories. Although we cannot establish whether such a mapping exists or not, a similar mapping for the vacua we have found in the finite volume analysis will not work, because expectation values of operators involving fermionic fields take a different value depending on the number of fermionic zero modes present. Notice that this result is the same for all values of $k$, the orbifold correspondence does not work even though orbifold symmetry is not broken. For instance, if we consider the operator 
\beq{eq:gauginocondensate}
\cO_p=\tr\left( \lambda^{\dag\,\alpha} \lambda^\alpha\right) \sim
\left({\it a}_a^{*\,\alpha}+ \sum_{n=1}^\infty {\it
  a}_a^{*\,\alpha}(n) e^{-2\pi i n x\over L}\right)\left({\it a}_a^{\alpha}+ \sum_{n=1}^\infty {\it a}_a^{\alpha}(n) e^{2\pi i n x\over L}\right)
\eeq
we are counting the number of fermions in the state, so there is a diagonal non-vanishing contribution
\beq{eq:numberferm}
\bra 0_B \right|\cU^l \cO_p \cU^{l'}\left| 0_B'\ket \sim 2l'\delta_{l,l'}\delta_{0_B,0_B'}
\eeq
However, the orbifold projection of this operator
\beq{eq:orbfcondensate}
\cO_d=\tr\left( \lambda_{12}^{\dag\,\alpha}\lambda_{12}^\alpha + \cdots + \lambda_{k1}^{\dag\,\alpha}\lambda_{k1}^\alpha\right)
\eeq
will have a vanishing diagonal contribution, since there are no fermionic zero modes over the vacuum state.

\subsection{Potential over moduli space.}\label{sec:moduli}

We now turn to the question of whether a potential appears over the Abelian moduli space of the daughter theory due to quantum effects. For simplicity, we will work with $U(1)$ fields $A_\mu = a_\mu \sqrt{1/N}{\bf 1}_N$. Gauge fields are uncharged, so they cannot give rise to a potential. Fermionic fields do couple through
\beq{eq:fermioncoupl}
\cL_{\lambda a} = \sum_{i=1}^k \sum_{a=1}^{N^2} {\overline \lambda}_{i,i+1}^a ({\partial\!\!\!\slash} +e^i{a\!\!\!\slash}^i-e^{i+1}{a\!\!\!\slash}^{i+1})\lambda_{i,i+1}^a 
\eeq
where $e^i$ is the Abelian coupling. In principle $e^i=e^{i+1}=e$, since all Abelian gauge groups enter symmetrically and we have applied the renormalization group from a point where all couplings were equal. At that point $e^2= g_t/N^2$, where $g_t=g_{YM}^2N$ is the non-Abelian 't Hooft coupling.

Let us comment several aspects of the Abelian moduli space. We can scale the fields so that the coupling constants come in front of the action. Then, when we make an Abelian gauge transformation that shifts the gauge field by a constant, the fermionic field charged under that Abelian group picks up a phase depending on the position. For instance, if 
\beq{eq:phase}
a_3^i-a_3^{i+1}\to a_3^i-a_3^{i+1}+c_3\ \ \Rightarrow \ \ \lambda_{i,i+1} \to e^{ic_3x_3}\lambda_{i,i+1}  
\eeq
then, when we move a period along the $x_3$ direction, the phase of the field changes by $e^{ic_3l_3}$. We conclude that the periods of the $\bS^1$ components of the Abelian moduli space are $2\pi/l_3$. Would be twisted gauge transformations correspond to $2\pi/N l_3$ translations along the $\bS^1$, although fixed boundary conditions for charged fields imply that they are no longer a symmetry of the theory.

Since there are no fermionic zero modes with support over the whole moduli space, we can integrate out the fermionic fields in the path integral. The regions of zero measure where fermionic zero modes have support will be localized at conical singularities of the effective potential obtained this way.
 
We proceed now to give the fermionic potential. At the one-loop level, it comes from the determinant that results in the path integral when we integrate out fermionic fields. 
\beq{eq:potferm}
V_{\rm eff}(a^i)=-\log\left( \prod_{i=1}^k \det\left( {\partial\!\!\!\!\slash} +{a\!\!\!\slash}^i-{a\!\!\!\slash}^{i+1}\right)^{N^2} \right)
\eeq
This potential can be computed using the methods of \cite{ref:luscher}. The result, after rescaling by the length of the spatial torus $a^i\to a^i/l$, is
\beq{eq:potferm2}
V_{\rm eff}(a^i) = -N^2 \sum_{i=1}^k V\left(a^i-a^{i+1}\right)
\eeq 
where the shape of the potential is given by
\beq{eq:su2pot}
V\left(a^i-a^{i+1}\right) = -{1\over \pi^2 l} \sum_{\bn\in \bZ^4} {\cos\left( n_\mu (a_\mu^i-a_\mu^{i+1})\right) \over (\bn^2)^2}
\eeq
Notice that the potential depends only on the differences $a^i-a^{i+1}$, so no potential is generated for the diagonal $U(1)$, only for the $U(1)^{k-1}$ groups coming from the non-Abelian part of the parent group after the orbifold projection. This potential has its minima at non-zero values of the differences, in the twisted sector. Since it is of order $N^2$, it cannot be ignored in the large $N$ limit. 

Examining more closely the potential, we can see that there is only one minimum at $a_\mu^i-a_\mu^{i+1}=\pi$. Usually, wavefunctions over the moduli space are characterized by the electric flux, that we can interpret as momentum along the moduli space. However, kinetic energy on the moduli space is proportional to $e^2$ (\ref{eq:hamabelian}), that is very small and becomes smaller as we increase the volume. On the other hand, the effective potential is very large and does not depend on the coupling. Therefore, the wavefunction is very localized, even for translations given by twisted transformations, and it is more convenient to use a position representation over the moduli space. The first consequence of all this is that if we make an Abelian twisted transformation, we are moving the system to a configuration with more energy, see Fig.~1. Therefore, most of the vacua of the theory are lifted. Only when we make diagonal non-Abelian twisted gauge transformations, the system remains in a ground state. This leaves $N$ vacua. 

\bfig{h!}{fig:abelianmoduli}
\bcen
\graphs{0.35}{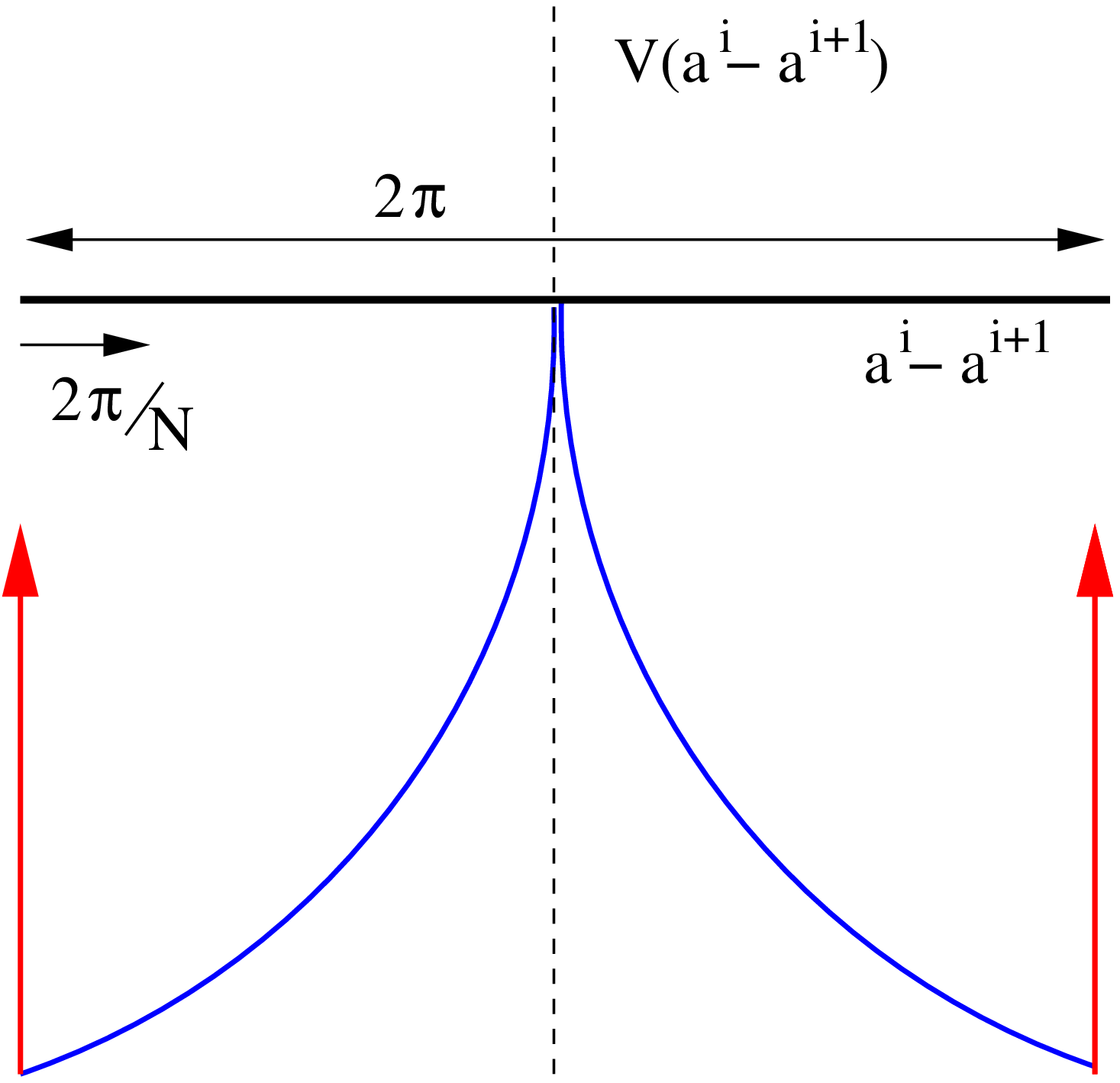}
\ecen 
\hspace{1.5cm}\parbox[c]{13cm}{\caption{\footnotesize The wavefunction of the ground state is localized at the minimum of the potential in the Abelian moduli space. A translation produced by a twisted transformation will lift the state to a configuration with more energy.}}
\efig

On the other hand, the wavefunctions over the moduli space of parent and daughter theories are very different. The first one can be seen as a zero-momentum state, while the last is more a position state. This suggests that in the infinite volume limit the daughter theory will fall into an Abelian twisted configuration, breaking orbifold symmetry in that sector. Another difference is that the ground state has a negative energy of order $O(N^2)$. These results are of the same kind as the ones found in \cite{ref:tong,ref:ourpaper}, although in the other cases the moduli space was non-Abelian for the daughter theory. 

We have found many differences between the vacua of parent and daughter theories in finite volume with maximal twist. We were looking to lift the non-Abelian moduli space that in previous works pointed towards non-perturbative failure of planar equivalence. We have seen that properly taking into account the Abelian groups in the daughter theory, the situation does not really improve. Thanks to Abelian groups we are able to map the moduli space of parent and daughter theories, but the generation of a potential by fermionic fields in the daughter theory makes the physical behavior of both theories quite different, although it lifts many of the unexpected non-Abelian vacua. The moduli space of the parent theory is also a problem, because it implies that there are fermionic zero modes that generate vacua that cannot be mapped to the daughter theory. 

\section{Tunneling effects.}

We have shown that parent and daughter theories have physically inequivalent vacua in this context due to the potential generated over the moduli space in the daughter theory and to the mismatch of vacua. However, some quantities do not depend on the moduli space, so the wavefunction will just give a normalization factor that can be chosen to be the same. The orbifold conjecture may still be useful to make some computations in the common vacua, up to kinematical factors. 

One of the main applications of twisted boundary conditions is the computation of fermion condensates \cite{ref:cesarcohen} generated by tunneling between different vacua. These fermion condensates do not depend on the moduli space and tunneling can be studied using self-dual solutions of Euclidean equations of motion. In the case where the tunneling is between vacua related by a twisted gauge transformation, the relevant configurations are of fractional topological charge, which we have associated to the twist tensor (\ref{eq:topcharge}). Notice that to map parent and daughter theories we are using (\ref{eq:twisttensors}) and (\ref{eq:conditions}). If we want to make the mapping between purely non-Abelian configurations, we should take $\alpha_{\mu\nu}=0$, otherwise we will be introducing self-dual Abelian fluxes in the daughter theory. In this case, all the twist tensors in the daughter theory must be equal because of fermions in the bifundamental representations. This implies that parent and daughter twist tensors are proportional 
\beq{eq:diagonaltwist}
n_{\mu\nu}^p=kn_{\mu\nu}^d
\eeq
Under these conditions, the fractional contribution to topological charge (\ref{eq:topcharge}) by tunneling is the same in both theories 
\beq{eq:topchargecompar}
\cQ = - {1\over kN} n_{\mu\nu}^p {\tilde{n}}_{\mu\nu}^p = k\left( -{1\over N} n_{\mu\nu}^d {\tilde n}_{\mu\nu}^d \right)
\eeq  

\subsection{Tunneling in parent theory.}\label{sec:parenttun}  

In the $SU(kN)$ parent theory with $n_{\mu\nu}^p=0\, {\rm mod}\, k$ twist, the solution of minimal charge has $\cQ= 1/N$, $k$ times the minimal possible topological charge of the theory. It contributes to matrix elements of the form $\bra 0\right| \cO {\hat U}\left|0\ket$ for some operator $\cO$. The operator ${\hat U}$ acts over the trivial gauge vacuum $\left|0\ket$ by making a twisted gauge transformation (\ref{eq:twistgaugecond}). We can compute this quantity using a Euclidean path integral
\beq{eq:matrixelement2}
\bra 0\right| \cO {\hat U}\left|0\ket = -\lim_{T\to \infty} {1\over T} \bra 0\right| \cO e^{-HT}{\hat U}\left|0\ket \to
\int_{\cQ=1/N} \cD A\cD\lambda \cD{\overline\lambda} \cO_E e^{-S_E}
\eeq
Although we do not know the explicit Euclidean solution that contributes in the saddle point approximation, we know what should be the vacuum expectation value of some operators. Because of supersymmetry, only zero modes contribute. There are $4k$ real zero modes of both bosonic and fermionic fields. The existence of fermionic zero modes implies that the tunneling contributes only to operators involving a product of $2k$ fermionic fields. In particular, there is no contribution to the vacuum energy. Therefore,
\beq{eq:energy}
\cE\equiv -\lim_{T\to\infty}{1\over T}\ln\bra 0\right|e^{-HT}{\hat U}\left|0\ket =0
\eeq
and
\beq{eq:fermionk}
\bra \tr_{kN}(\lambda\lambda)^k\ket\equiv -\lim_{T\to\infty}{1\over T}\bra 0\right| \tr_{kN}(\lambda\lambda)^k e^{-HT}{\hat U}\left|0 \ket = {1\over k!}\left(\bra \tr_{kN} \lambda\lambda\ket\right)^k
\eeq
which is a tunneling contribution to a $2k$-fermionic correlation function. By $\bra \tr_{kN} \lambda\lambda\ket$ we refer to the value of the gaugino condensate in the parent theory. This does not mean that a gaugino condensate is generated by these configurations, we are just referring to the numerical value of (\ref{eq:fermionk}). This is based on the use of torons to estimate the gaugino condensate \cite{ref:cesarcohen} and on the coincidence of the instanton measure with the measure of a superposition of torons \cite{ref:zhit}. The factorial factor comes from considering the configuration of charge $\cQ=1/N$ as a superposition of $k$ equal configurations of charge $\cQ=1/kN$. 
 
Another peculiar fact about this configuration is that it is transformed into itself under a Nahm transformation \cite{ref:nahm}. A Nahm transformation identifies moduli spaces of different self-dual configurations of different gauge theories in different spaces. From a stringy perspective it is a remnant of T-duality \cite{ref:nahmtdual}.
   
We will use techniques developed in \cite{ref:nahmtorus,ref:nahmtw} for Nahm transformations with twisted boundary conditions. Given the rank $r$, the topological charge $\cQ$, and the twist 
\beq{eq:nahmtensor}
n=\left( \begin{array}{cc} 0  & \Xi\\ -\Xi & 0 \end{array}\right)
\eeq
where $\Xi=\,{\rm diag}\,(q_1,q_2)$, the Nahm transformed quantities are given by
\beq{eq:nahmtransform}
\begin{array}{rclcrcl}
p_1 &  = & N/\,{\rm gcd}\,(q_1,N) & \ \ & p_2 &  = & N/\,{\rm gcd}\,(q_2,N) \\  s_1q_1& = & -\,{\rm gcd}\,(q_1,N)\,{\rm mod}\, N & \ & s_2q_2 &= &-\,{\rm gcd}\,(q_2,N)\,{\rm mod}\, N\\
\cQ' & = & r/p_1p_2 & \ \ & r' & = & \cQ p_1p_2\\
\end{array}
\eeq 
and the twist
\beq{eq:nahmtransform2}
\Xi'=\,{\rm diag}\,\left((p_1-s_1)p_2\cQ, (p_2-s_2)p_1\cQ \right).
\eeq

Our tunneling configuration can be determined by $\vecm=(0,-k,0)$, $\veck=(0,1,0)$ for instance. Then, we can write the twist tensor as $\Xi=\,{\rm diag}\,(1,k)$. Using (\ref{eq:nahmtransform}) and (\ref{eq:nahmtransform2}), it is straightforward to see that $r'=r=kN$, $\cQ'=\cQ=1/N$ and $\Xi'=\Xi=\,{\rm diag}\,(1,k)$.

\subsection{Tunneling in daughter theory.}\label{sec:daughtertun}

Self-dual solutions contributing to tunneling in the daughter theory do not coincide with the projection from the parent theory, as given by (\ref{eq:diagonaltwist}). Tunneling between two adjacent vacua is given by a configuration where each gauge factor contributes by $\cQ_d=1/N$ to the topological charge. Then, the total topological charge is $\cQ=k/N$, while the topological charge in the parent theory is $\cQ_p=1/N$, in disagreement with (\ref{eq:topchargecompar}).

The configurations we are using for tunneling are not related by the map we have proposed, but notice that we are comparing semiclassical contributions of the same order, since the classical Euclidean action in parent $S_{cl}^p =8\pi^2/g_p^2N$ and daughter $S_{cl}^d = 8\pi^2k/g_d^2N$ theories have the same value according to the orbifold projection $kg_p^2=g_d^2$. When we examine the moduli space of parent and daughter tunneling configurations, we find further evidence that we are comparing the correct quantities. The bosonic moduli space is $4k$-dimensional for both theories, and the moduli space of the daughter theory also transforms into itself under a Nahm transformation. This can be seen using (\ref{eq:nahmtransform}) for each of the gauge factors: $\cQ_d'=\cQ_d=1/N$, $r_d'=r_d=N$, $\Xi_d'=\Xi_d=\,{\rm diag}\,(1,1)$.

Even using this 'improved' mapping, disagreement emerges from fermionic fields. In the daughter theory there is no supersymmetry in general, because fermions are in a representation different to bosons. However, bifundamental representations of the daughter theory can be embedded in the adjoint representation of the  parent theory, and this will be useful.

In order to make an explicit computation, we need a formula for the self-dual gauge configuration of fractional topological charge. We do not have an analytic expression in general, but 't Hooft found a set of solutions \cite{ref:torons}, called torons, that are self-dual when the sizes of the torus $l_\mu$ satisfy some relations. We will assume that we are in the good case and that, although in other cases the relevant configurations will be different, the physics will be the same.

 Torons are the solutions of minimal topological charge in a situation with maximal twist. For a $SU(N)$ gauge theory, this means that $\cQ_{\rm toron}=1/N$. Therefore, they can be used in the daughter theory, where we will have a toron for each of the non-Abelian groups. We can compare torons to the most known self-dual solutions, instantons. Instanton solutions can be characterized by their size, orientation in the algebra and center position. The size of torons is fixed by the size of the spatial torus, and they have a fixed orientation along a $U(1)$ subgroup. However, they still have a center that can be put at any point of the torus. Therefore, the moduli space of torons is a four-dimensional torus. An explicit expression for torons in the daughter theory in the presence of magnetic flux in the $x_3$ direction is 
\beq{eq:toron}
A_\mu^i={i\pi\over 2gl_\mu l_\nu}\eta_{\mu\nu}^3 (x-z^i)_\nu\,\omega \ , \ i=1,\dots,k
\eeq 
where $\eta_{\mu\nu}^a$, $a=1,2,3$ are 't Hooft's self-dual eta symbols \cite{ref:instantons}, $\omega$ is a generator of a $U(1)\subset SU(N)$ and $z^i$ are the center positions. Notice that we have chosen to work with anti-hermitian and canonically normalized gauge fields. 

Now that we are armed with an analytic expression, we can calculate what happens with fermions in the background of $k$ torons. The Euclidean Dirac operator is
\beq{eq:diracop}
\gamma_\mu D_\mu = \left(\begin{array}{cc} 0 & -i\sigma_\mu D_\mu \\  i{\overline\sigma}_\mu D_\mu& 0\end{array}\right)
\eeq
where $\sigma_\mu= ({\bf 1}_2, i\vec{\sigma})$, ${\overline \sigma}_\mu=\sigma_\mu^\dag$ and $\sigma_i$ are the Pauli matrices. In a self-dual background $F_{\mu\nu}^+$, the square of the Dirac operator is positive definite for negative chirality fields
\beq{eq:rightdiracop}
(i{\overline\sigma}_\mu D_\mu)(i\sigma_\nu D_\nu)=-D^2{\bf 1}_2
\eeq
while for positive chirality fields, it has a potentially negative contribution
\beq{eq:leftdiracop}
(i\sigma_\mu D_\mu)(i{\overline\sigma}_\nu D_\nu)=-D^2{\bf 1}_2 - {1\over 2}
g\sigma_{\mu\nu}F_{\mu\nu}^+
\eeq
where $\sigma_{\mu\nu}= \sigma_{[\mu}{\overline \sigma}_{\nu]}$. This means that there are no fermionic zero modes of negative chirality, although there can be zero modes of positive chirality. In the supersymmetric case, where fermions are in the adjoint representation, there are two zero modes in the background of a single toron.

The covariant derivative acting over bifundamental fields is
\beq{eq:covdev}
D_\mu\lambda_{i,i+1}=\partial_\mu\lambda_{i,i+1}+gA_\mu^i\lambda_{i,i+1}-g\lambda_{i,i+1}A_\mu^{i+1}
\eeq
Then, introducing (\ref{eq:toron}) and (\ref{eq:covdev}) in (\ref{eq:leftdiracop}), we find the operator
\beq{eq:diracorbf}
\left(-D^2{\bf 1}_2 -{1\over 2}g\sigma_{\mu\nu}F_{\mu\nu}^+\right)_{\rm adj}-2g\Delta A_\mu^i (D_\mu)_{\rm adj}{\bf 1}_2+M^2(\Delta z)\omega^2\otimes {\bf 1}_2
\eeq
where ``adj'' denotes that the operator is equal to the case where it acts over the adjoint representation of $SU(N)$. Extra contributions are given by the separation between torons of different groups
\beq{eq:deltas}
\Delta z_\mu^i=z_\mu^{i+1}-z_\mu^i \ \ , \ \ \Delta A_\mu^i= {i\pi \over 2gl_\mu l_\nu}\eta_{\mu\nu}^3 \Delta z_\nu^i\,\omega
\eeq
 The coefficient of the positive `mass' contribution is
\beq{eq:mass}
M^2(\Delta z)={\pi^2\over 4} \left({(\Delta z_0^i)^2+(\Delta z_3^i)^2\over l_0^2l_3^2}+{(\Delta z_1^i)^2+(\Delta z_2^i)^2\over l_1^2l_2^2}\right)
\eeq
The physics of fermions is clear now. When two torons of adjacent groups coincide at the same space-time point, there are zero modes for bifundamental fermions transforming under these groups. Zero modes are identical to the case of adjoint representation (supersymmetric case). When we separate the torons, the zero modes acquire a mass proportional to the distance of separation. 

Therefore, we can split the moduli space of torons $\cM_k$ in sectors where a different number of torons are coincident. We can use quiver diagrams to represent different sectors. Each node is a toron position. Links joining different nodes represent massive fermions, while links coming back to the same node represent fermions effectively in the adjoint representation, so they give rise to supersymmetric contributions. The expansion will look like\beq{eq:expansion}
\begin{array}{rccccc}
\cM_k  & = &
\begin{array}{c}
\graphs{0.45}{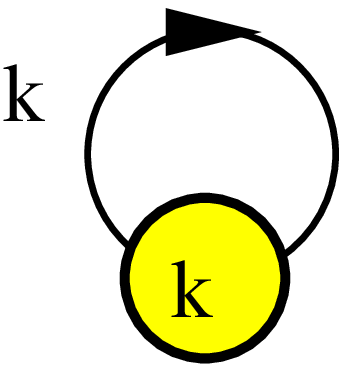}
\end{array}
& +& 
\begin{array}{c}
\graphs{0.45}{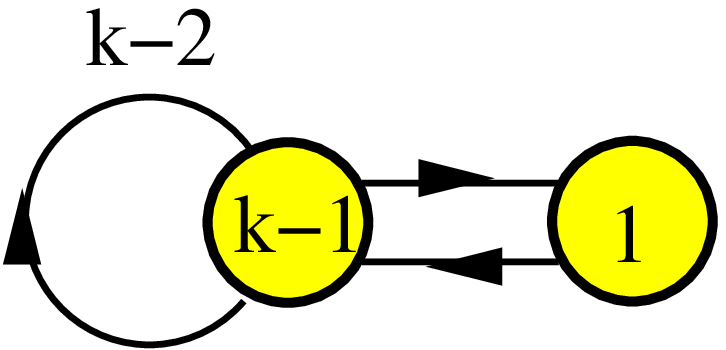}
\end{array}
& +\\ & +& 
\begin{array}{c}
\graphs{0.5}{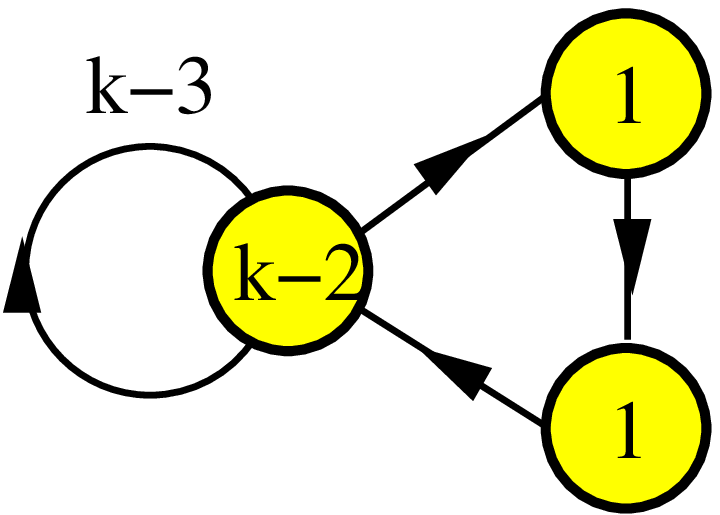}
\end{array}
& +& \cdots&  
\end{array}
\eeq
A sector with $n$ adjoint links can contribute only to vacuum expectation values of quantities involving at least $n$ factors of the form $\lambda_{i,i+1}\lambda_{i,i+1}$. The contribution is determined by the expectation value of the gaugino condensate of a $\cN=1$ $SU(N)$ supersymmetric gauge theory, $\bra \lambda\lambda\ket_{\rm susy}$. As an example, consider the pure supersymmetric contribution
\beq{eq:expansionterms1}
\begin{array}{rcl}
\left(\begin{array}{c}
\graphs{0.45}{quiver1.eps}
\end{array}\right)_{\prod_{i=1}^k\tr_N\lambda_{i,i+1}\lambda_{i,i+1}} & = & \left( \bra \lambda\lambda\ket_{\rm susy} \right)^k
\end{array}
\eeq
that receives non-supersymmetric corrections from other sectors of the moduli space, for instance
\beq{eq:expansionterms2}
\begin{array}{l}
\left(\begin{array}{c}
\graphs{0.45}{quiver2.eps}
\end{array}\right)_{\prod_{i=1}^k\tr_N\lambda_{i,i+1}\lambda_{i,i+1}} =  \\ \\  =  k\left( \bra \lambda\lambda\ket_{\rm susy} \right)^{k-2} \int dA_1(z^1)dA_2(z^2)\, {\tilde M}^4(z^1-z^2)\,e^{-2S_{cl}}\, \bra \tr{1\over {D\!\!\!\!\slash}_{12}}\tr{1\over {D\!\!\!\!\slash}_{21}} \ket' 
\end{array}
\eeq
% {\det'({D\!\!\!\!\slash}_{12})\det'({D\!\!\!\!\slash}_{21})\over (\det'(-D^2_1))^{1/2}(\det'(-D^2_2))^{1/2}}
we denote by $A_i$ the bosonic zero modes, depending on the toron position $z^i$. The factors ${\tilde M}^2$ come from fermionic would-be zero modes, and make the contribution to vanish in the regions of moduli space where the separated toron coincide with the others. We have taken care of bosonic and fermionic (would be) zero modes, so they have been subtracted from the computation of the correlation function involving the fermion propagators $\bra \tr {D\!\!\!\!\slash}_{12}^{-1} \tr {D\!\!\!\!\slash}_{21}^{-1}\ket'$. The classical action is the action of the toron solution
\beq{eq:claction}
S_{cl}={8\pi^2\over g^2 N}
\eeq

Coming back to (\ref{eq:energy}), we see that there is a non-zero contribution of tunneling to energy. When two torons are coincident, there are fermionic zero modes, so contributions to $\cE$ come from the sector of the moduli space of torons where all are separated. This contribution is completely non-supersymmetric
\beq{eq:energycont}
\!\!\!\!\cE_{\rm 1-loop} = \int \prod_{i=1}^k dA_i(z^i) {\tilde M}^2(z^i-z^{i+1})e^{-kS_{cl}}{\prod_{i=1}^k
  \det'({D\!\!\!\!\slash}_{i,i+1}(A_i,A_{i+1}))\over \prod_{i=1}^k
  (\det'(-D^2_i(A_i)))^{1/2}}  
\eeq
We have taken care of bosonic and fermionic (would be) zero modes, so they have been subtracted from the determinants $\det'$. With the notation we are employing, the contribution to $\cE$ will come from a ``ring'' diagram Fig.~2.
\bfig{h!}{fig:quiver}
\bcen
\graphs{0.5}{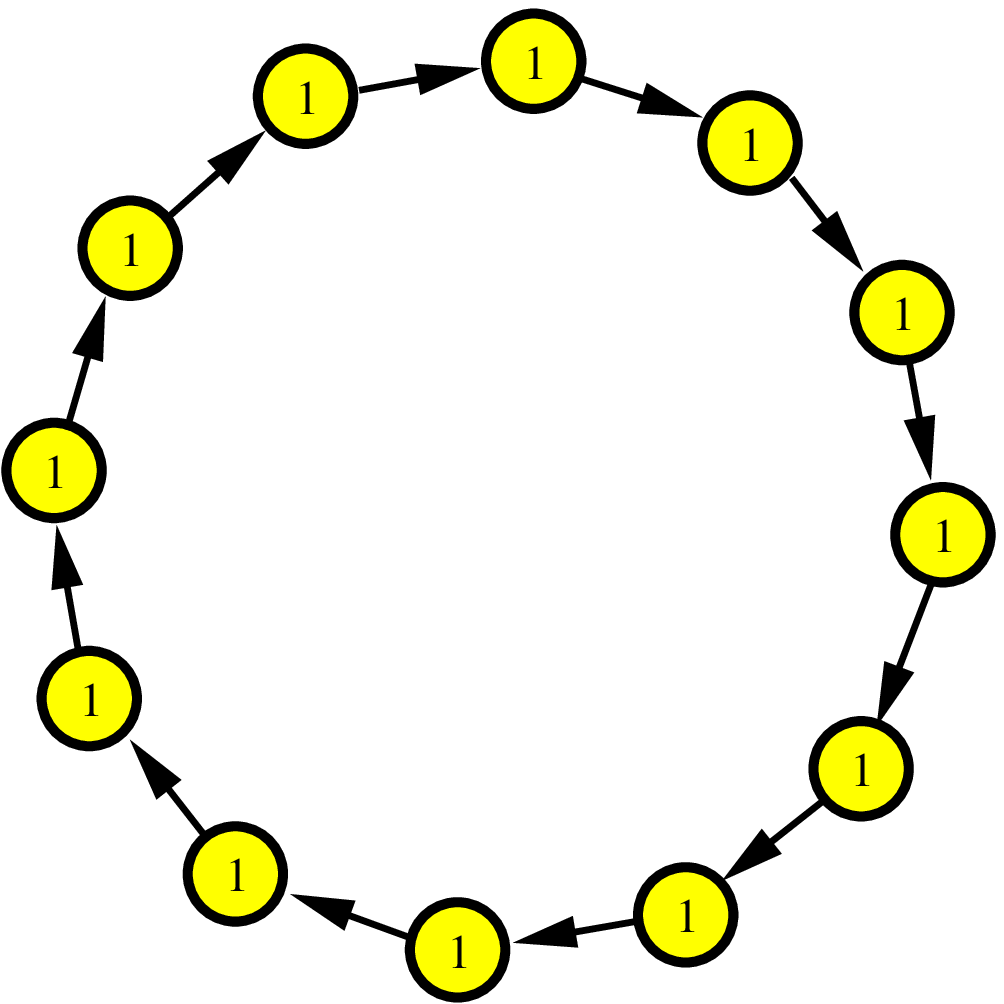}
\ecen 
\caption{\footnotesize The ring sector produces no supersymmetric factors.}
\efig

When we make the orbifold projection of $(\ref{eq:fermionk})$, we find 
\beq{eq:fermionkorbf}
\tr_{kN} (\lambda\lambda)^k \to \tr_N
(\lambda_1\cdots\lambda_k)^2= \left( \bra\lambda\lambda\ket_{\rm susy}\right)^k+k\left(\bra\lambda\lambda\ket_{\rm susy}\right)^{k-2}F^{(1)}_{NS}+\dots
\eeq
Where the first non-supersymmetric factor $F_{NS}^{(1)}$ is the same as in (\ref{eq:expansionterms2}), and the dots indicate that there are more contributions from other sectors of moduli space, all involving different non-supersymmetric factors. Notice that the orbifold mapping $k g_p^2= g_d$, implies that the renormalization invariant scales of parent and daughter theories are the same at the planar level. 
\beq{eq:invscales}
\bra \tr_{kN}\lambda\lambda\ket \sim  \Lambda_p^3 =\mu^3\exp\left({-8\pi^2\over g_p^2 kN}\right) = \mu^3\exp\left({-8\pi^2\over g_d^2 N}\right) =\Lambda_d^3 \sim \bra\lambda\lambda\ket_{\rm susy}
\eeq
As a consequence, the first term in (\ref{eq:fermionkorbf}) coincides with (\ref{eq:fermionk}), up to the factorial factor, that in the daughter theory does not appear because torons belonging to different groups are distinguishable.

So we must conclude that there are no simple relations between (\ref{eq:energy}) and (\ref{eq:energycont}) or between (\ref{eq:fermionk}) and (\ref{eq:fermionkorbf}) although both sets of quantities are invariant under orbifold symmetry.

\subsection{Semiclassical dependence on the vacuum angles.}

The results of the previous sections can be used to analyze the dependence on the vacuum angles when we deform the theories adding a mass $m$ for the fermions. This is necessary in order to have such a dependence, otherwise massless fermions will erase it through the chiral anomaly. Notice that we can make the fermions of the daughter theory massive only for the $\bZ_2$ orbifold, so we will restrict to this case. 

In the parent theory, when we introduce a mass for the fermions, the states previously associated to fermionic zero modes (\ref{eq:fermionvacua}) are lifted by an energy that is roughly the mass times the number of fermionic modes. The energy density is obtained dividing by the volume of the box $V=l^3$. So we have $k=2$ levels, each with $N$ states labelled by the electric flux. As we will see in a moment, tunneling between states of the same level produces a lifting in the energy density of order $m^4$. If we are in the small volume (small mass) limit $ml\ll 1$, then the lifting produced by fermionic modes is larger than tunneling contributions, so the states in the highest level decouple. However, as we increase the volume there will be level-crossings between states of different levels.  

In the daughter theory, a mass for the fermions modifies the effective potential over the Abelian moduli space (\ref{eq:su2pot}). If $ml\ll 1$, then only high frequency contributions are suppressed, but increasing the volume will exponentially suppress the whole potential, so non-diagonal vacua could become relevant.  

We are interested in estimating the dependence of the vacuum energy on the vacuum angles and electric fluxes
\beq{eq:vacenergytheta}
\cE(e,\theta)= \lim_{T\to\infty} -{1\over T}\ln \bra e,\theta\right| e^{-HT} \left| e,\theta\ket.
\eeq
In a semiclassical expansion, the relevant contributions come from fractional instantons that encode tunneling processes between different vacua. We will first study the leading term given by a single (anti)toron of minimal topological charge, and afterwards we will estimate the contribution of infinite many (anti)torons using a dilute gas approximation \cite{ref:coleman}.

In the parent theory, the lowest order contribution is  
$$
\bra 0_p \right| e^{-HT} {\hat U}_p\left| 0_p \ket =T \int d A(z^1,z^2) m^4 \exp\left(-{8\pi^2\over g_p^2N} \right) {\det_F'\left(iD\!\!\!\!\slash + m\right) \over \det_B' \left( -D^2\right)^{1/2}}
$$
\beq{eq:parenttoron}
=T m^4 K_p(m) e^{-2 S_{cl}}
\eeq  
where $A(z^1,z^2)$ are the zero modes of the $SU(2N)$ gauge field, depending on eight parameters of the $\cQ=1/N$ toron configuration. The fermionic mass is $m$, so the factor $m^4$ comes from the would be fermionic zero modes that appear when we turn $m$ to zero. Therefore, they have been subtracted from the determinants $\det_F'$. The bosonic determinant $\det_B'$ includes only physical polarizations of non-zero modes of vector bosons, so any ghost contribution has been taken into account in it. In the last equality, we have used the orbifold map $g_p^2 = g_d^2/2$. $K_p(m)$ is a $m$-dependent constant that will be obtained after making the integral over the moduli space of the (anti)toron.

For a general transition amplitude between different twisted vacua, there can be contributions of an arbitrary number of torons and anti-torons, as long as the total topological charge has the correct value. Since we are in a dilute gas approximation, we neglect possible interactions,
\beq{eq:parentgeneral}
\bra 0_p\right| ({\hat U}_p^+)^{l_+} e^{-HT} ({\hat U}_p)^{l_-}\left|0_p\ket = \sum_{n=0}^\infty \sum_{\nbar=0}^\infty {1\over n! \nbar!}\left(T m^4 K_p(m) e^{-2S_{cl}}\right)^{n+\nbar}\delta_{n-\nbar-l_++l_-}. 
\eeq  
We are ready now to compute the vacuum energy in the state (\ref{eq:ekstates}), it is straightforward to see that
\beq{eq:parentenergy}
\cE_p(e_p,\theta_p)= -2 m^4 K_p(m) e^{-2 S_{cl}} \cos\left[{2\pi\over N}\left(e_p+{\theta_p\over 2\pi}\right) \right] 
\eeq

A similar analysis can be done for the daughter theory. The lowest order contribution is, in the diagonal case,
$$
\bra 0_d \right| e^{-HT} {\hat U}_1{\hat U}_2\left| 0_d \ket =T \int d A_1(z^1) d A_2(z^2) (m+{\tilde M}(\Delta z))^4 e^{-2S_{cl}} {\det_F'\left(-D_{12}^2 + m^2\right)\over \det_B' \left( -D_1^2\right)^{1/2}\det_B' \left( -D_2^2\right)^{1/2}}
$$
\beq{eq:daughtertoron}
=T m^4 K_d(m) e^{-2 S_{cl}}
\eeq 
We have argued above that it is plausible that the bosonic measures of parent $d A(z^1,z^2)$ and daughter $dA_1(z^1)dA_2(z^2)$ are the same. However, there are extra contributions to the $m^4$ factor, given by the mass $M(\Delta z)$ (\ref{eq:mass}) that bifundamental modes acquire when the two torons are separated. We can also see that the one-loop determinants in (\ref{eq:parenttoron}) and (\ref{eq:daughtertoron}) are different. All this imply that $K_p(m)$ and $K_d(m)$ are not simply related, as we would have expected from the orbifold projection.

Then, a general transition between diagonal vacua is given by
\beq{eq:daughtergeneral}
\bra 0_d\right| ({\hat U}_2^+{\hat U}_1^+)^{l_+} e^{-HT} ({\hat U}_1{\hat U}_2)^{l_-}\left|0_d\ket = \sum_{n=0}^\infty \sum_{\nbar=0}^\infty  {1\over n! \nbar!}\left(T m^4 K_d(m) e^{-2S_{cl}}\right)^{n+\nbar}\delta_{n-\nbar-l_++l_-}. 
\eeq  
So the vacuum energy of the state (\ref{eq:elstates}) will be 
\beq{eq:daughterenergy}
\cE_d^D(e_1,e_2,\theta_1,\theta_2)= -2 m^4 K_d(m) e^{-2 S_{cl}} \cos\left[{2\pi\over N}\left(e_1+e_2+{\theta_1+\theta_2\over 2\pi}\right) \right] 
\eeq

\bfig{h!}{fig:spectralflow}
\bcen
\graph{10cm}{8cm}{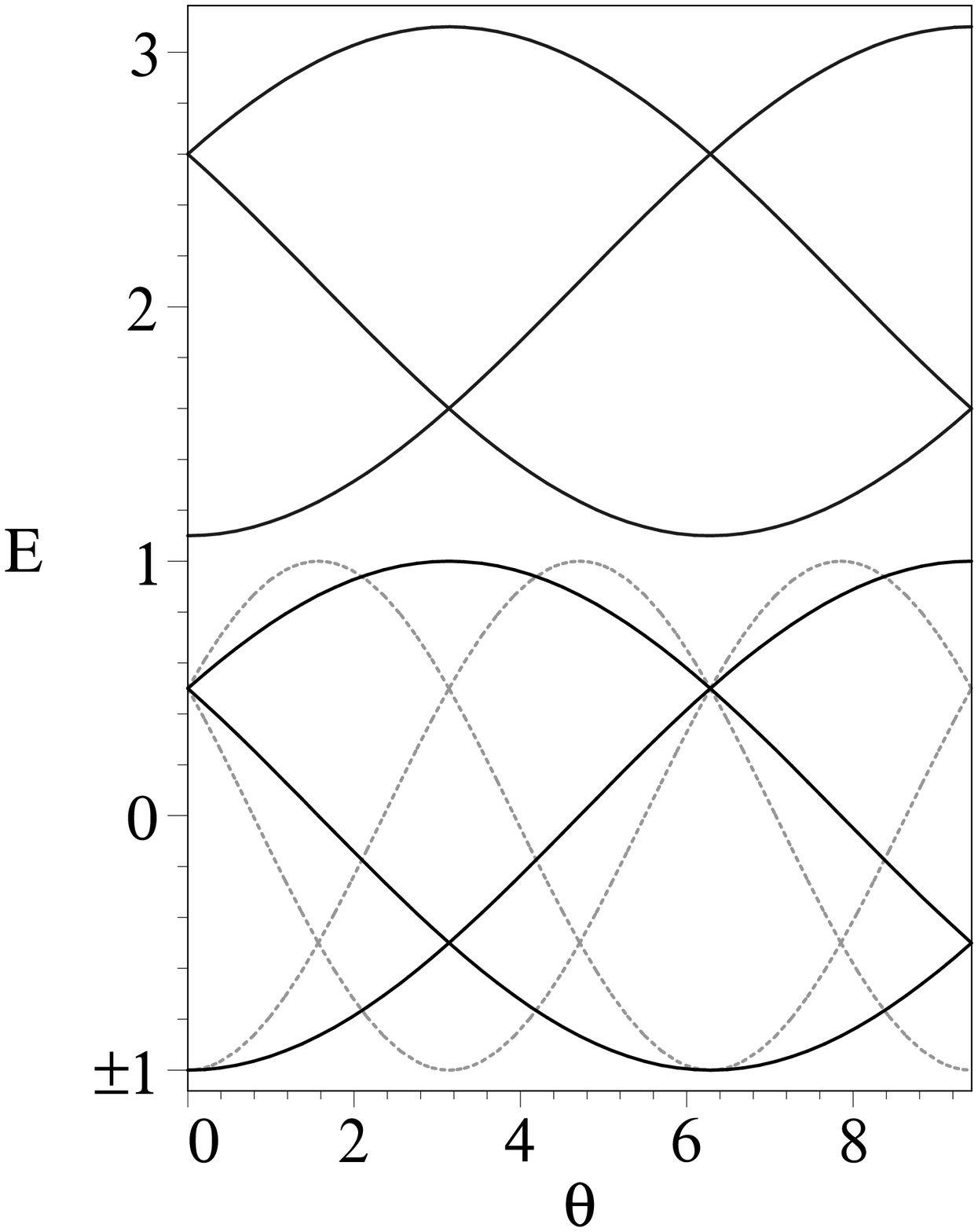}
\ecen 
\caption{\footnotesize We show a picture of the spectral flow of $\cE_p/K_p$ (dark lines) and $\cE_d/K_d$ (light lines) for different states of electric flux in the case $N=3$. The horizontal axis corresponds to $\theta_p$ for the parent theory and to $\theta_d$ for the daughter theory. We assume $ml\ll 1$, so the parent theory presents two separated levels. Notice that the periodicity of the spectral flow of the daughter is half the periodicity of the parent.}
\efig

We can study the spectral flows induced by a change in the vacuum angles \cite{ref:vanbaal} in both theories and compare the results. In Fig.~3, we illustrate the discussion with a simple example. In the parent theory, the energy of a state of definite electric flux is invariant only under shifts $\theta_p\to\theta_p+2\pi N$. However, the whole spectrum is invariant under a $2\pi$ rotation of $\theta_p$, so there is a non-trivial spectral flow that may become non-analytic and produce oblique confinement in the infinite volume limit. Notice also that there are two well-separated levels in the small volume ($ml \ll 1$) limit, where the lifted states have fermionic constant Abelian modes. In the opposite limit, states of the same electric flux number become nearly degenerate. This could be a hint for a two-to-one mapping between parent and daughter states of equivalent electric flux, in a finite volume version of the statement made in \cite{ref:yaffetheta}. Notice that the parent and daughter ``vacuum angles'' should be mapped as $2\theta_d=\theta_p$ at the orbifold point $\theta_1=\theta_2=\theta_d$, as the orbifold correspondence dictates. Indeed, we find that the spectrum of the daughter theory is invariant under $\theta_d\to\theta_d+\pi$. However, this does not look as the right transformation for a vacuum angle and, in view of (\ref{eq:daughterenergy}), the daughter vacuum angle seems rather to be $\theta_1+\theta_2$, that has the value $2\theta_d$ at the orbifold point. 

On the other hand, the lifted vacua of the parent theory are unsuitable for a correspondence with daughter states because some matrix elements involving fermions will be different, as we have argued before. Regarding the remaining vacua, there is a quantitative disagreement between parent (\ref{eq:parentenergy}) and daughter energies (\ref{eq:daughterenergy}), given by the difference in the values of the tunneling coefficients $K_p(m)$ and $K_d(m)$, so it seems that planar equivalence does not work at this level. 

\section{D-brane interpretation.}

We can give a geometrical interpretation of our results based on a construction with D-branes in the torus, as in \cite{ref:dtorons}. In the parent theory we will have a set of $kN$ D4-branes on $\bT^4$, the 01234 directions including time. The low energy theory is a $U(kN)$ supersymmetric gauge theory. We are interested in study possible vacuum configurations, that are characterized by different bundles of D-branes over the torus. We will ignore the scalars associated to the transverse dimensions to the torus, anyway, they will be spectators in the analysis below. 

The twist of the configuration associated to tunneling in the parent theory, Sec.~\ref{sec:parenttun}, is realized diluting D2-brane charge in the worldvolume of D4s. The reason are the couplings of D2 charge with $U(1)$ first Chern class of D4 gauge group
\beq{eq:d2coupl}
\int_{D4} C_3 \wedge \tr F 
\eeq
the twist induced in the $U(1)$ must be compensated by the non-Abelian part, in order to have a well-defined $U(kN)$ bundle. So we can introduce $k$ units of magnetic flux in the 3 direction and one unit of electric flux in the same direction by introducing $k$ D2 branes in 34 direction and a D2 brane in 12 directions. Since the intersection number is non-zero we can interpret also this configuration as having D2 at some angle inside the torus.
\bcen
\begin{tabular}{|l|c|c|c|c|c|}
\hline
     & 0 & 1 & 2 & 3 & 4  \\\hline
kN D4& $\times$ & $\times$ & $\times$ & $\times$ & $\times$ \\ \hline
k  D2& $\times$ & $\bullet$ & $\bullet$ & $\times$ & $\times$  \\\hline
1  D2& $\times$ & $\times$ & $\times$ & $\bullet$ & $\bullet$ \\ \hline 
\end{tabular}
\ecen
After making a T-duality along 12 directions 
\bcen
\begin{tabular}{|l|c|c|c|c|c|}
\hline
     & 0 & 1 & 2 & 3 & 4 \\\hline
kN D2& $\times$ & $\bullet$ & $\bullet$ & $\times$ & $\times$ \\\hline
k  D4& $\times$ & $\times$ & $\times$ & $\times$ & $\times$ \\\hline
1  D0& $\times$ & $\bullet$ & $\bullet$ & $\bullet$ & $\bullet$  \\ \hline
\end{tabular}
\ecen
this configuration is equivalent to an instanton for a $U(k)$ theory with $kN$ units of magnetic flux. However, if the T-duality transformation is made along 34 directions 
\bcen
\begin{tabular}{|l|c|c|c|c|c|}
\hline
     & 0 & 1 & 2 & 3 & 4 \\\hline
kN D2& $\times$ & $\times$ & $\times$ & $\bullet$ & $\bullet$ \\\hline
k  D0& $\times$ & $\bullet$ & $\bullet$ & $\bullet$ & $\bullet$ \\\hline
1  D4& $\times$ & $\times$ & $\times$ & $\times$ & $\times$ \\ \hline
\end{tabular}
\ecen
we will have $k$ `instantons' in a $U(1)$ theory with $kN$ units of electric flux.

In the daughter theory, we have a set of $N$ D4 branes wrapped around a $\bT^4$ but sitting at an orbifold point in transverse space. The twist associated to the tunneling configuration in the daughter theory, Sec.~\ref{sec:daughtertun}, is realized diluting a D2-brane both in 34 and 12 directions. The orbifold action is responsible of the multiples copies of the gauge group. So physically we are introducing what in an unorbifolded theory will be $N$ D4 branes and not $kN$ branes, although the field content is constructed projecting the last.
\bcen
\begin{tabular}{|l|c|c|c|c|c|}
\hline
     & 0 & 1 & 2 & 3 & 4  \\\hline
(k)N  D4& $\times$ & $\times$ & $\times$ & $\times$ & $\times$  \\\hline
(k)1  D2& $\times$ & $\bullet$ & $\bullet$ & $\times$ & $\times$  \\\hline
(k)1  D2& $\times$ & $\times$ & $\times$ & $\bullet$ & $\bullet$  \\ \hline
\end{tabular}
\ecen
We can make a T-duality transformation along 12 directions 
\bcen
\begin{tabular}{|l|c|c|c|c|c|}
\hline
     & 0 & 1 & 2 & 3 & 4\\ \hline
(k)N  D2& $\times$ & $\bullet$ & $\bullet$ & $\times$ & $\times$ \\\hline
(k)1  D4& $\times$ & $\times$ & $\times$ & $\times$ & $\times$ \\\hline
(k)1  D0& $\times$ & $\bullet$ & $\bullet$ & $\bullet$ & $\bullet$ \\ \hline
\end{tabular}
\ecen
so we have an `instanton' in the presence of $(k)N$ units of magnetic flux in a $U(1)^k$ theory. If the T-duality transformation is made along 34 directions 
\bcen
\begin{tabular}{|l|c|c|c|c|c|}
\hline
     & 0 & 1 & 2 & 3 & 4 \\\hline
(k)N  D2& $\times$ & $\times$ & $\times$ & $\bullet$ & $\bullet$ \\\hline
(k)1  D0& $\times$ & $\bullet$ & $\bullet$ & $\bullet$ & $\bullet$ \\\hline
(k)1  D4& $\times$ & $\times$ & $\times$ & $\times$ & $\times$ \\ \hline
\end{tabular}
\ecen
now the flux is electric instead of magnetic. In this two cases, we can interpret the `untwisted' fractional instanton as a D0 brane in the presence of a D2 brane. This is the sector where all torons are at the same point. The sectors where torons are at different points correspond to the splitting of the D0 in fractional D0s that can move independently along the torus directions. The gauge group is the same for $k$ fractional D0s as for a regular D0, however the strings joining different fractional D0s acquire a mass proportional to the separation. This is reflected in the mass we have computed for fermionic modes (\ref{eq:mass}). So in fact, the quiver diagrams we have used to label different sectors of the moduli space correspond to the gauge theory living in the D0s at different points of the moduli space.

From this point of view we also have a geometric picture of why the configuration of the daughter theory should be mapped to a configuration of the parent theory with $k$ units of both electric and magnetic flux. The matching is between the number of fractional branes in the orbifolded theory and the number of branes of the same dimensions in the `parent'. 

If we make another T-duality transformation along 1234 directions,
\bcen
\begin{tabular}{|l|c|c|c|c|c|}
\hline
     & 0 & 1 & 2 & 3 & 4 \\\hline
(k)N  D0& $\times$ & $\bullet$ & $\bullet$ & $\bullet$ & $\bullet$ \\\hline
(k)1  D2& $\times$ & $\times$ & $\times$ & $\bullet$ & $\bullet$ \\\hline
(k)1  D2& $\times$ & $\bullet$ & $\bullet$ & $\times$ & $\times$ \\ \hline
\end{tabular}
\ecen
the D4s become D0s that can move along the torus. The separation in fractional D0s is encoded in the Abelian part of the daughter theory. The diagonal $U(1)$ correspond to the center of mass position, while the $U(1)^{k-1}$ group that form the Abelian moduli space are the relative separations of D0s in this T-dual interpretation. From (\ref{eq:potferm2}) we know that D0s tend to separate, a signal of the orbifold tachyon.

\section{Summary.}

We extend the study of planar equivalence in a small volume \cite{ref:ourpaper} introducing twisted boundary conditions for orbifold field theories. We have found several sources of disagreement with orbifold large $N$ equivalence. First of all, a `kinematical' map seems not to work properly. When we try to embed the vacuum of the daughter theory in the vacuum of the parent theory, we find difficulties because the number of vacua does not match due to the presence (absence) of fermionic zero modes in the parent (daughter) theory. We also find difficulties in identifying the relevant configurations for tunneling using only algebraic arguments. 

An important `dynamical' indication of disagreement with planar equivalence is the formation of a potential over the Abelian moduli space of the daughter theory, which makes the ground wavefunction of both theories very different and shifts the vacuum energy at leading order. These results are analogous to the behavior found for the non-Abelian moduli space in the case of periodic boundary conditions \cite{ref:ourpaper}.

When we try to compute contributions produced by tunneling effects in the daughter theory, we find a remarkable relation with a supersymmetric theory. We can split the contributions in sectors where some factors are identical to quantities of a supersymmetric theory (to be precise, to the gaugino condensate). However, the relation is with a $\cN=1$ $SU(N)$ gauge theory (or $k$ copies of it) and not with the parent theory, that has group $SU(kN)$. We also compute the semiclassical dependence of the energy on the electric flux for the case $k=2$, but we fail in establishing a quantitative mapping between states of parent and daughter theories.

In view of these results, it seems that the `supersymmetric' properties of the orbifold daughter stem from the fact that a $\cN=1$ $SU(N)$ supersymmetric sector is `wrapping' it, the sector of diagonal configurations. This is in agreement with previous results \cite{ref:ourpaper} and suggests that planar equivalence of orbifold theories could be traced to the fact that the parent theory and $\cN=1$ $SU(N)$ gauge theory are also planar equivalent \cite{ref:shifmanlast}. From the point of view of finite volume physics, the fact that the orbifold theory can be embedded in a supersymmetric theory, the parent theory, apparently has no more meaning except that it is probably a necessary condition for having a supersymmetric sector. It would be interesting to check whether the properties of the orientifold theory relies on having a $SO(N)$ supersymmetric sector or work differently to the orbifold case.  

\section*{Acknowledgments.}
I would like to thank J.L.F.Barb\'on for reading the manuscript and making many comments and suggestions and for many useful discussions. This work was supported by Spanish MEC FPU grant AP2002-0433.

%%%%%%%%%%%%%%%%%%%%%%%%%%%%%%%%%%%  BIBLIOGRAPHY  %%%%%%%%%%%%%%%%%%%%%%%

%%%%%%%%%%%%%%%%%%%%%%%% LIST OF FIGURES
%\listoffigures

%%%%%%%%%%%%%%%%%%%%%%%%%%%%%%%% DOCUMENT ENDS

\begin{thebibliography}{999}
%\bibitem[label]{marker}, \cite{marker}

%\bibitem{ref:susys}
%N.~Seiberg and E.~Witten,
  %``Electric - magnetic duality, monopole condensation, and confinement in N=2
  %supersymmetric Yang-Mills theory,''
%  Nucl.\ Phys.\ B {\bf 426} (1994) 19
%  [Erratum-ibid.\ B {\bf 430} (1994) 485]
%  [arXiv:hep-th/9407087].
  %%CITATION = HEP-TH 9407087;%%
%N.~Seiberg and E.~Witten,
  %``Monopoles, duality and chiral symmetry breaking in N=2 supersymmetric
  %QCD,''
%  Nucl.\ Phys.\ B {\bf 431} (1994) 484
%  [arXiv:hep-th/9408099].
  %%CITATION = HEP-TH 9408099;%%
%N.~Seiberg,
  %``Electric - magnetic duality in supersymmetric nonAbelian gauge theories,''
%  Nucl.\ Phys.\ B {\bf 435} (1995) 129
%  [arXiv:hep-th/9411149].
  %%CITATION = HEP-TH 9411149;%%

%\bibitem{ref:susys2}
%M.~A.~Shifman and A.~I.~Vainshtein,
%  %``Instantons versus supersymmetry: Fifteen years later,''
%  arXiv:hep-th/9902018.
%  %%CITATION = HEP-TH 9902018;%%

\bibitem{ref:thooftlargen}
 G.~'t Hooft,
  %``A Planar Diagram Theory For Strong Interactions,''
  Nucl.\ Phys.\ B {\bf 72} (1974) 461.
  %%CITATION = NUPHA,B72,461;%%

\bibitem{ref:strassler}
M.~J.~Strassler,
  %``On methods for extracting exact non-perturbative results in
  %non-supersymmetric gauge theories,''
  arXiv:hep-th/0104032.
  %%CITATION = HEP-TH 0104032;%%

%\bibitem{ref:orbfst}
% M.~R.~Douglas and G.~W.~Moore,
%  %``D-branes, Quivers, and ALE Instantons,''
%  arXiv:hep-th/9603167.
%  %%CITATION = HEP-TH 9603167;%%

%\bibitem{ref:orifst}
% E.~G.~Gimon and J.~Polchinski,
%  %``Consistency Conditions for Orientifolds and D-Manifolds,''
%  Phys.\ Rev.\ D {\bf 54} (1996) 1667
%  [arXiv:hep-th/9601038].
  %%CITATION = HEP-TH 9601038;%%

\bibitem{ref:orbifold}
%Mogoll del orbifold
 S.~Kachru and E.~Silverstein,
  %``4d conformal theories and strings on orbifolds,''
  Phys.\ Rev.\ Lett.\  {\bf 80} (1998) 4855
  [arXiv:hep-th/9802183].
  %%CITATION = HEP-TH 9802183;%%
 A.~E.~Lawrence, N.~Nekrasov and C.~Vafa,
  %``On conformal field theories in four dimensions,''
  Nucl.\ Phys.\ B {\bf 533}, 199 (1998)
  [arXiv:hep-th/9803015].
  %%CITATION = HEP-TH 9803015;%%
M.~Bershadsky, Z.~Kakushadze and C.~Vafa,
  %``String expansion as large N expansion of gauge theories,''
  Nucl.\ Phys.\ B {\bf 523}, 59 (1998)
  [arXiv:hep-th/9803076].
  %%CITATION = HEP-TH 9803076;%%
M.~Bershadsky and A.~Johansen,
  %``Large N limit of orbifold field theories,''
  Nucl.\ Phys.\ B {\bf 536}, 141 (1998)
  [arXiv:hep-th/9803249].
  %%CITATION = HEP-TH 9803249;%%
M.~Schmaltz,
  %``Duality of non-supersymmetric large N gauge theories,''
  Phys.\ Rev.\ D {\bf 59}, 105018 (1999)
  [arXiv:hep-th/9805218].
  %%CITATION = HEP-TH 9805218;%%

\bibitem{ref:orientifold}
%Mogollon del orientifold
Z.~Kakushadze,
  %``Gauge theories from orientifolds and large N limit,''
  Nucl.\ Phys.\ B {\bf 529}, 157 (1998)
  [arXiv:hep-th/9803214].
  %%CITATION = HEP-TH 9803214;%%
Z.~Kakushadze,
  %``On large N gauge theories from orientifolds,''
  Phys.\ Rev.\ D {\bf 58}, 106003 (1998)
  [arXiv:hep-th/9804184].
  %%CITATION = HEP-TH 9804184;%%
 F.~Sannino and M.~Shifman,
  %``Effective Lagrangians for orientifold theories,''
  Phys.\ Rev.\ D {\bf 69}, 125004 (2004)
  [arXiv:hep-th/0309252].
  %%CITATION = HEP-TH 0309252;%%
A.~Armoni, M.~Shifman and G.~Veneziano,
  %``From super-Yang-Mills theory to QCD: Planar equivalence and its
  %implications,''
  arXiv:hep-th/0403071.
  %%CITATION = HEP-TH 0403071;%%
A.~Armoni, A.~Gorsky and M.~Shifman,
  %``An exact relation for N = 1 orientifold field theories with arbitrary
  %superpotential,''
  Nucl.\ Phys.\ B {\bf 702}, 37 (2004)
  [arXiv:hep-th/0404247].
  %%CITATION = HEP-TH 0404247;%%
  P.~Di Vecchia, A.~Liccardo, R.~Marotta and F.~Pezzella,
  %``The gauge / gravity correspondence for non-supersymmetric theories,''
  Fortsch.\ Phys.\  {\bf 53}, 450 (2005)
  [arXiv:hep-th/0412234].
  %%CITATION = HEP-TH 0412234;%%
F.~Sannino,
  %``Higher representations: Confinement and large N,''
  arXiv:hep-th/0507251.
  %%CITATION = HEP-TH 0507251;%%

\bibitem{ref:largenequiv}
%Yaffe et al., el paper de estados coherentes y orbifold
P.~Kovtun, M.~Unsal and L.~G.~Yaffe,
  %``Necessary and sufficient conditions for non-perturbative equivalences of
  %large N(c) orbifold gauge theories,''
  JHEP {\bf 0507} (2005) 008
  [arXiv:hep-th/0411177].
  %%CITATION = HEP-TH 0411177;%%

\bibitem{ref:orientifoldproof}
A.~Armoni, M.~Shifman and G.~Veneziano,
  %``Exact results in non-supersymmetric large N orientifold field theories,''
  Nucl.\ Phys.\ B {\bf 667}, 170 (2003)
  [arXiv:hep-th/0302163].
  %%CITATION = HEP-TH 0302163;%%
A.~Armoni, M.~Shifman and G.~Veneziano,
  %``Refining the proof of planar equivalence,''
  Phys.\ Rev.\ D {\bf 71} (2005) 045015
  [arXiv:hep-th/0412203].
  %%CITATION = HEP-TH 0412203;%%

\bibitem{ref:quarkcond}
A.~Armoni, M.~Shifman and G.~Veneziano,
  %``SUSY relics in one-flavor QCD from a new 1/N expansion,''
  Phys.\ Rev.\ Lett.\  {\bf 91}, 191601 (2003)
  [arXiv:hep-th/0307097].
  %%CITATION = HEP-TH 0307097;%%
  A.~Armoni, M.~Shifman and G.~Veneziano,
  %``QCD quark condensate from SUSY and the orientifold large-N expansion,''
  Phys.\ Lett.\ B {\bf 579}, 384 (2004)
  [arXiv:hep-th/0309013].
  %%CITATION = HEP-TH 0309013;%%

\bibitem{ref:newresults1}
 A.~Armoni and E.~Imeroni,
  %``Predictions for orientifold field theories from type 0' string theory,''
  arXiv:hep-th/0508107.
  %%CITATION = HEP-TH 0508107;%%

\bibitem{ref:newresults2}
 T.~A.~Ryttov and F.~Sannino,
  %``Hidden QCD in chiral gauge theories,''
  arXiv:hep-th/0509130.
  %%CITATION = HEP-TH 0509130;%%

\bibitem{ref:forandagainst}
  A.~Gorsky and M.~Shifman,
  %``Testing nonperturbative orbifold conjecture,''
  Phys.\ Rev.\ D {\bf 67}, 022003 (2003)
  [arXiv:hep-th/0208073].
  %%CITATION = HEP-TH 0208073;%%
 R.~Dijkgraaf, A.~Neitzke and C.~Vafa,
  %``Large N strong coupling dynamics in non-supersymmetric orbifold field
  %theories,''
  arXiv:hep-th/0211194.
  %%CITATION = HEP-TH 0211194;%%
J.~Erlich and A.~Naqvi,
  %``Nonperturbative tests of the parent/orbifold correspondence in
  %supersymmetric gauge theories,''
  JHEP {\bf 0212}, 047 (2002)
  [arXiv:hep-th/9808026].
  %%CITATION = HEP-TH 9808026;%%
 P.~Kovtun, M.~Unsal and L.~G.~Yaffe,
  %``Non-perturbative equivalences among large N(c) gauge theories with  adjoint
  %and bifundamental matter fields,''
  JHEP {\bf 0312}, 034 (2003)
  [arXiv:hep-th/0311098].
  %%CITATION = HEP-TH 0311098;%%
%En el que dan muchos argumentos en contra del orbifold
 A.~Armoni, A.~Gorsky and M.~Shifman,
  %``Spontaneous Z(2) symmetry breaking in the orbifold daughter of N = 1
  %super-Yang-Mills theory, fractional domain walls and vacuum structure,''
  Phys.\ Rev.\ D {\bf 72}, 105001 (2005)
  [arXiv:hep-th/0505022].
  %%CITATION = HEP-TH 0505022;%%

\bibitem{ref:yaffetheta}
%Yaffe, donde se discute multivaluacion de vacios
P.~Kovtun, M.~Unsal and L.~G.~Yaffe,
  %``Can large N(c) equivalence between supersymmetric Yang-Mills theory and its
  %orbifold projections be valid?,''
  arXiv:hep-th/0505075.
  %%CITATION = HEP-TH 0505075;%%

\bibitem{ref:shifmanlast}
%Shifman (et al?), el ultimo review, o donde primero se mencione que el condensado no sirve como order parameter
M.~Shifman,
  %``Non-perturbative Yang-Mills from supersymmetry and strings, or, in the
  %jungles of strong coupling,''
  arXiv:hep-th/0510169.
  %%CITATION = HEP-TH 0510169;%%

\bibitem{ref:tong}
%el de Tong
D.~Tong,
  %``Comments on condensates in non-supersymmetric orbifold field theories,''
  JHEP {\bf 0303}, 022 (2003)
  [arXiv:hep-th/0212235].
  %%CITATION = HEP-TH 0212235;%%

\bibitem{ref:ourpaper}
%nuestro paper del potencial
J.~L.~F.~Barbon and C.~Hoyos,
  %``Small volume expansion of almost supersymmetric large N theories,''
  arXiv:hep-th/0507267.
  %%CITATION = HEP-TH 0507267;%%

\bibitem{ref:emflux}
%'t Hooft, electric and magnetic flux in non-Abelian gauge theories
 G.~'t Hooft,
  %``A Property Of Electric And Magnetic Flux In Nonabelian Gauge Theories,''
  Nucl.\ Phys.\ B {\bf 153} (1979) 141.
  %%CITATION = NUPHA,B153,141;%%

\bibitem{ref:index}
%Witten, susy index
E.~Witten,
  %``Constraints On Supersymmetry Breaking,''
  Nucl.\ Phys.\ B {\bf 202}, 253 (1982).
  %%CITATION = NUPHA,B202,253;%%

\bibitem{ref:index2}
%Witten, susy index in 4d
E.~Witten,
  %``Supersymmetric index in four-dimensional gauge theories,''
  Adv.\ Theor.\ Math.\ Phys.\  {\bf 5}, 841 (2002)
  [arXiv:hep-th/0006010].
  %%CITATION = HEP-TH 0006010;%%

\bibitem{ref:ymtorus}
%Toni, YM on 4d torus: classical theory
A.~Gonzalez-Arroyo,
  %``Yang-Mills fields on the 4-dimensional torus. (Classical theory),''
  arXiv:hep-th/9807108.
  %%CITATION = HEP-TH 9807108;%%

\bibitem{ref:torons}
%'t Hooft, twisted self-dual solutions
G.~'t Hooft,
  %``Some Twisted Selfdual Solutions For The Yang-Mills Equations On A
  %Hypertorus,''
  Commun.\ Math.\ Phys.\  {\bf 81}, 267 (1981).
  %%CITATION = CMPHA,81,267;%%

\bibitem{ref:cita27toni}
%cita 27 del review de toni
D.~R.~Lebedev, M.~I.~Polikarpov and A.~A.~Rosly,
  %``Summation Over Topological Classes Of Gauge Fields In Lattice Gauge
  %Theories,''
  Sov.\ J.\ Nucl.\ Phys.\  {\bf 49} (1989) 1113
  [Yad.\ Fiz.\  {\bf 49} (1989) 1799].
  %%CITATION = SJNCA,49,1113;%%

\bibitem{ref:luscher}
%Luscher, donde calcula el potencial a un loop en el toro
M.~Luscher,
  %``Some Analytic Results Concerning The Mass Spectrum Of Yang-Mills Gauge
  %Theories On A Torus,''
  Nucl.\ Phys.\ B {\bf 219} (1983) 233.

\bibitem{ref:cesarcohen}
 E.~Cohen and C.~Gomez,
  %``Chiral Symmetry Breaking In Supersymmetric Yang-Mills,''
  Phys.\ Rev.\ Lett.\  {\bf 52} (1984) 237.
  %%CITATION = PRLTA,52,237;%%
  E.~Cohen and C.~Gomez,
  %``A Computation Of Tr(-1)**F In Supersymmetric Gauge Theories With Matter,''
  Nucl.\ Phys.\ B {\bf 223}, 183 (1983).
  %%CITATION = NUPHA,B223,183;%%

\bibitem{ref:zhit}
  A.~R.~Zhitnitsky,
  %``Torons, Chiral Symmetry Breaking And The U(1) Problem In The Sigma Model
  %And In Gauge Theories,''
  Nucl.\ Phys.\ B {\bf 340} (1990) 56.
  %%CITATION = NUPHA,B340,56;%%

\bibitem{ref:nahm}
 M.~F.~Atiyah, N.~J.~Hitchin, V.~G.~Drinfeld and Y.~I.~Manin,
  %``Construction Of Instantons,''
  Phys.\ Lett.\ A {\bf 65} (1978) 185.
  %%CITATION = PHLTA,A65,185;%%
%transformaciones de Nahm
W.~Nahm,
  %``A Simple Formalism For The Bps Monopole,''
  Phys.\ Lett.\ B {\bf 90}, 413 (1980).
  %%CITATION = PHLTA,B90,413;%%
P.~J.~Braam and P.~van Baal,
  %``Nahm's Transformation For Instantons,''
  Commun.\ Math.\ Phys.\  {\bf 122} (1989) 267.
  %%CITATION = CMPHA,122,267;%%

\bibitem{ref:nahmtdual}
 E.~Witten,
  %``Sigma models and the ADHM construction of instantons,''
  J.\ Geom.\ Phys.\  {\bf 15} (1995) 215
  [arXiv:hep-th/9410052].
  %%CITATION = HEP-TH 9410052;%%
  E.~Witten,
  %``Small Instantons in String Theory,''
  Nucl.\ Phys.\ B {\bf 460} (1996) 541
  [arXiv:hep-th/9511030].
  %%CITATION = HEP-TH 9511030;%%
%Nahm transformation y T-duality
D.~E.~Diaconescu,
  %``D-branes, monopoles and Nahm equations,''
  Nucl.\ Phys.\ B {\bf 503}, 220 (1997)
  [arXiv:hep-th/9608163].
  %%CITATION = HEP-TH 9608163;%%
 A.~Astashkevich, N.~Nekrasov and A.~Schwarz,
  %``On noncommutative Nahm transform,''
  Commun.\ Math.\ Phys.\  {\bf 211} (2000) 167
  [arXiv:hep-th/9810147].
  %%CITATION = HEP-TH 9810147;%%
 M.~Hamanaka and H.~Kajiura,
  %``Gauge fields on tori and T-duality,''
  Phys.\ Lett.\ B {\bf 551}, 360 (2003)
  [arXiv:hep-th/0208059].
  %%CITATION = HEP-TH 0208059;%%

\bibitem{ref:nahmtorus}
 H.~Schenk,
  %``On A Generalized Fourier Transform Of Instantons Over Flat Tori,''
  Commun.\ Math.\ Phys.\  {\bf 116} (1988) 177.
  %%CITATION = CMPHA,116,177;%%
A.~Gonzalez-Arroyo and C.~Pena,
  %``Nahm transformation on the lattice,''
  JHEP {\bf 9809} (1998) 013
  [arXiv:hep-th/9807172].
  %%CITATION = HEP-TH 9807172;%%
P.~van Baal,
  %``Nahm gauge fields for the torus,''
  Phys.\ Lett.\ B {\bf 448} (1999) 26
  [arXiv:hep-th/9811112].
  %%CITATION = HEP-TH 9811112;%%
 M.~Garcia Perez, A.~Gonzalez-Arroyo, C.~Pena and P.~van Baal,
  %``Nahm dualities on the torus: A synthesis,''
  Nucl.\ Phys.\ B {\bf 564} (2000) 159
  [arXiv:hep-th/9905138].
  %%CITATION = HEP-TH 9905138;%%

\bibitem{ref:nahmtw}
%Toni , transformaciones de Nahm twisted
 A.~Gonzalez-Arroyo,
  %``On Nahm's transformation with twisted boundary conditions,''
  Nucl.\ Phys.\ B {\bf 548}, 626 (1999)
  [arXiv:hep-th/9811041].
  %%CITATION = HEP-TH 9811041;%%

\bibitem{ref:instantons}
%instantones de 't Hooft
 % A.~A.~Belavin, A.~M.~Polyakov, A.~S.~Shvarts and Y.~S.~Tyupkin,
  %``Pseudoparticle Solutions Of The Yang-Mills Equations,''
 % Phys.\ Lett.\ B {\bf 59}, 85 (1975).
  %%CITATION = PHLTA,B59,85;%%
G.~'t Hooft,
  %``Computation Of The Quantum Effects Due To A Four-Dimensional
  %Pseudoparticle,''
  Phys.\ Rev.\ D {\bf 14}, 3432 (1976)
  [Erratum-ibid.\ D {\bf 18}, 2199 (1978)].
  %%CITATION = PHRVA,D14,3432;%%

\bibitem{ref:coleman}
S.~Coleman, ``Aspects of symmetry: selected Erice Lectures.'', Cambridge University Press, 1985.

\bibitem{ref:vanbaal}
P.~van Baal, ``Twisted Boundary Conditions: a Non-Perturbative Probe for Pure Non-Abelian Gauge Theories.'', PhD Thesis (Utrecht, 1984).
P.~van Baal,
  %``QCD in a finite volume,''
  arXiv:hep-ph/0008206.

\bibitem{ref:dtorons}
%torones vistos como estados bound the d-branas
 Z.~Guralnik and S.~Ramgoolam,
  %``Torons and D-brane bound states,''
  Nucl.\ Phys.\ B {\bf 499} (1997) 241
  [arXiv:hep-th/9702099].
  %%CITATION = HEP-TH 9702099;%%
Z.~Guralnik and S.~Ramgoolam,
  %``From 0-branes to torons,''
  Nucl.\ Phys.\ B {\bf 521} (1998) 129
  [arXiv:hep-th/9708089].
  %%CITATION = HEP-TH 9708089;%%

\end{thebibliography}
\end{document}